%% file: main.tex
\documentclass[sigconf,screen,nonacm]{acmart}
\usepackage{popets}

\usepackage{amsfonts}
\usepackage{amsmath}
\usepackage{booktabs}
\usepackage{caption}
\usepackage{comment}
\usepackage{diagbox}
\usepackage{enumitem}
\usepackage{graphicx}
\usepackage{hyperref}
\usepackage{multirow}
\usepackage{pifont}
\usepackage{subfigure}
\usepackage{xcolor}

\AtBeginDocument{%
  }
\DeclareMathOperator*{\argmax}{arg\,max}

\setcopyright{popets}
\copyrightyear{YYYY}
\acmYear{YYYY}
\acmVolume{YYYY}
\acmNumber{X}
\acmDOI{XXXXXXX.XXXXXXX}
\acmISBN{}
\acmConference{Proceedings on Privacy Enhancing Technologies}
\begin{document}

\title{Investigating the Effect of Misalignment on Membership Privacy in the White-box Setting}

\author{Ana-Maria Cretu}
\authornote{To appear in the Proceedings on Privacy Enhancing Technologies (PoPETs 2024).}
\authornote{Work done partially at Microsoft as part of an internship and partially at ICL.}
\affiliation{
  \institution{EPFL}
  \city{}\country{}
  }
\email{ana-maria.cretu@epfl.ch}

\author{Daniel Jones}
\affiliation{
  \institution{M365 Research}
    \city{}\country{}
  }
\email{jonesdaniel@microsoft.com}

\author{Yves-Alexandre de Montjoye}
\affiliation{%
  \institution{Imperial College London (ICL)}
    \city{}\country{}
  }
\email{deMontjoye@imperial.ac.uk}

\author{Shruti Tople}
\affiliation{%
  \institution{Azure Research}
    \city{}\country{}
  }
\email{shruti.tople@microsoft.com}

\renewcommand{\shortauthors}{Ana-Maria Cre\c tu, Daniel Jones, Yves-Alexandre de Montjoye and Shruti Tople}

\input{sections/abstract}

\maketitle

\input{sections/introduction}

\input{sections/background}

\input{sections/related_work}

\input{sections/causes_misalignment}

\input{sections/alignment}

\input{sections/membership_inference}

\input{sections/discussion}

\input{sections/acknowledgments}

\bibliographystyle{ACM-Reference-Format}
\bibliography{bibliography}

\input{sections/appendix}

\end{document}

%% file: sections/abstract.tex
\begin{abstract}
Machine learning models have been shown to leak sensitive information about their training datasets. 
Models are increasingly deployed on devices, 
raising concerns that white-box access to the model parameters increases the attack surface compared to black-box access which only provides query access.
Directly extending the shadow modelling technique from the black-box to the white-box setting has been shown, in general, not to perform better than black-box only attacks. 
A potential reason is misalignment, a known characteristic of deep neural networks. 
In the shadow modelling context, misalignment means that, while the shadow models learn similar features in each layer, the features are located in different positions. 
We here present the first systematic analysis of the causes of misalignment in shadow models and show the use of a different weight initialisation to be the main cause. 
We then extend several re-alignment techniques, previously developed in the model fusion literature, to the shadow modelling context, where the goal is to re-align the layers of a shadow model to those of the target model.
We show re-alignment techniques to significantly reduce the measured misalignment between the target and shadow models. 
Finally, we perform a comprehensive evaluation of white-box membership inference attacks (MIA). Our analysis reveals that internal layer activation-based MIAs suffer strongly from shadow model misalignment, while gradient-based MIAs are only sometimes significantly affected. 
We show that re-aligning the shadow models strongly improves the former's performance and can also improve the latter's performance, although less frequently.
On the CIFAR10 dataset with a false positive rate of 1\%, white-box MIA using re-aligned shadow models improves the true positive rate by 4.5\%.
Taken together, our results highlight that on-device deployment increases the attack surface and that the newly available information can be used to build more powerful attacks.\footnote{Source code available at \url{https://github.com/microsoft/shadow-realignment-mia}.}
\end{abstract}

%% file: sections/introduction.tex
\section{Introduction}
\label{sec:introduction}
Machine learning (ML) models are being increasingly adopted by businesses, governments, and organisations.
The datasets they are trained on often contain information about individuals, such as pictures, documents, and metadata.

Inference attacks have been used to empirically measure the privacy risks of ML models.
Running an inference attack involves, first, defining a secret which an adversary aims to infer from a model, e.g., whether a target record was used to train the model, referred to as a membership inference attack (MIA). 
Second, it involves learning a distinguishing boundary between models having different values of the secret.
Using MIA as an example, the adversary aims to learn a good rule for distinguishing models trained on the target record from models trained without it~\cite{carlini2022membership}.
The main approach for learning this boundary relies on so-called shadow models~\cite{ateniese2015hacking,shokri2017membership,carlini2022membership}.
Shadow models are trained by an adversary using the same architecture as the target model, on similar datasets, while controlling the value of the secret in each dataset.
After training shadow models, the adversary extracts their weights~\cite{ganju2018property}, or queries the models on well-chosen records, to extract features such as gradients and activations of different layers~\cite{shokri2017membership,nasr2019comprehensive,zhang2021leakage}.
These pairs of (1) a model's behavior (as captured through a set of well-chosen features) and (2) the corresponding value of the secret, can be used to train a meta-classifier for inferring the secret (e.g., a membership classifier).

Shadow modelling has powered a broad range of inference attacks against ML models, including privacy attacks aiming to infer the membership of a record~\cite{shokri2017membership,salem2018ml,choquette2021label,carlini2022membership} or to reconstruct a record~\cite{salem2020updates,balle2022reconstructing}, property inference attacks aiming to infer global properties of the dataset~\cite{ateniese2015hacking,zhang2021leakage,suri2022formalizing}, and correlation inference attacks aiming to infer  correlation coefficients between input columns~\cite{crectu2021correlation}.

In spite of all this research, ML privacy risks have mostly been studied in the black-box setting, where the adversary accesses the model through an API~\cite{dionysiou2023sok,niu2023sok}.
This threat model is inadequate for measuring privacy risks of on-device and open-source models.
In particular, we are seeing an increasing trend in deploying models on user devices (edge machine learning).
This trend is motivated by a need on the deployers' end to reduce bandwidth and latency in serving requests, as well as user privacy expectations.
For instance, users' data should not leave their device or be accessed by third parties~\cite{applecsam}, and personalised models, e.g., through fine-tuning on the user's private data, should similarly not leave the users' devices. 

\begin{figure*}[!ht]
    \centering
    \subfigure[Input image
    ]{
        \centering
       \includegraphics[scale=0.2]{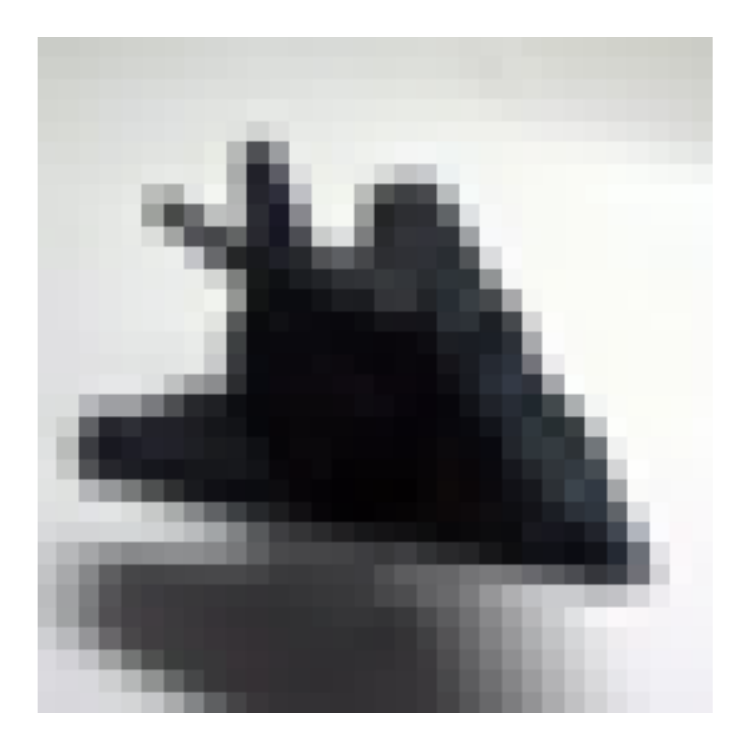} 
    }
    \hfil\\
    \subfigure[Activations of a target model $\mathcal{M}_T$]{
        \includegraphics[width=0.8\linewidth]{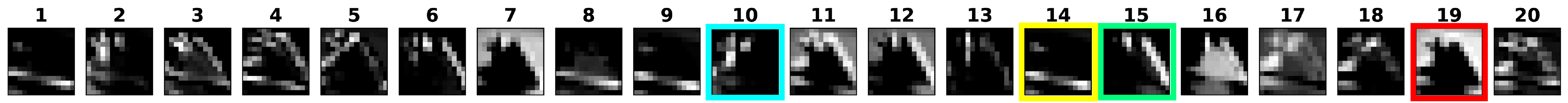}
    }
    \hfil
    \subfigure[Activations of shadow models (one model/row) trained by the classical adversary]{
            \includegraphics[width=0.8\linewidth]{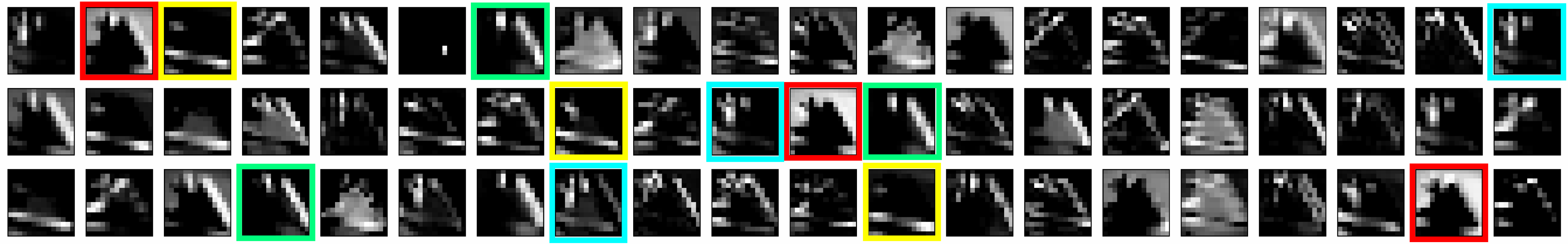}
    }
    \hfil
    \subfigure[Activations of re-aligned shadow models]{
        \includegraphics[width=0.8\linewidth]{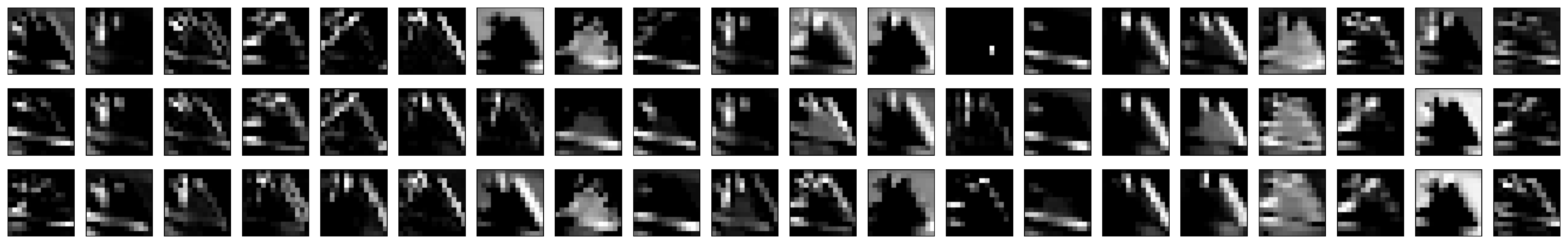}
    }
    \caption{Visualisation of the misalignment issue in CNN shadow models trained by the classical adversary, where models learn similar features (marked by the same color) located in different positions (c). We also visualise the same models after re-alignment (d). We refer the reader to Sec.~\ref{subsec:misalignment} for a detailed description of the figure. }\label{fig:visualise_misalignment}
\end{figure*}

Measuring the risk \textit{only through the lens of black-box adversaries} can lead to underestimating it~\cite{suri2022formalizing}.
Indeed, adversaries targeting open-source or on-device models have access to the model architecture and weights, which are not available in black-box access. 
Open-source models are usually released as files storing the weights (called a checkpoint) with code to load the weights in a class implementing the model~\cite{huggingface}.
Models can also be extracted from devices, e.g., Deng et al.~\cite{deng2022understanding} extracted 245 models from 62.5K Android apps.
While accessing and analysing app files is more difficult in iOS compared to Android, Hu et al.~\cite{hu2023first} demonstrated that even models deployed in iOS apps can be extracted.
Another example showing that motivated actors can reverse-engineer on-device models is Apple's NeuralHash model~\cite{applecsam}.
Shortly after Apple announced its proposed system design to detect child sexual abuse images on devices before users would upload them to iCloud, the model was reverse-engineered and the code allowing anyone to do this was made public~\cite{techcrunch2021}.
Furthermore, once a model is released publicly or on edge devices, it cannot be taken back, and stronger attacks can be mounted at a later date. Thus, developing powerful white-box attacks is extremely important.

However, it has been shown that the \textit{na\"ive} extension of shadow modelling to the white-box setting, by using features of internal layers (such as activations or weights), 
leads to worse attack performance than in the black-box setting~\cite{nasr2019comprehensive,zhang2021leakage,crectu2021correlation}.
Combining features of internal layers with model confidences (already available in the black-box setting) does not necessarily lead to better performance compared to the black-box setting~\cite{song2021systematic,crectu2021correlation,liu2022ml}.

In this work, we focus on standard deep neural network (DNN) architectures such as convolutional neural networks (CNN) and multilayer perceptrons and investigate shadow model misalignment as a potential reason for the sub-optimal performance of white-box MIAs.
Our starting point is that the adversary must train shadow models on a different dataset (otherwise they would already have perfect knowledge of the target model's training dataset) and must use a different randomness.
Due to DNN symmetries, such as the permutation equivalence property~\cite{ashmore2015method,ganju2018property} where the neurons of internal layers can be arbitrarily permuted without changing the network's function, shadow models trained on different data and randomness end up \textit{misaligned}. 
More specifically, although they learn similar features in the same layer, the features are not located in the same position (see Fig.~\ref{fig:visualise_misalignment} for an illustration in CNN shadow models).
Ganju et al.~\cite{ganju2018property} showed misalignment to affect the performance of white-box property inference attacks, arguing that the meta-classifier needs to learn the symmetries of its input features in addition to the inference task. 
However, no work so far has investigated the  causes of misalignment in shadow models nor quantitatively measured its impact on MIAs.

\textbf{Contribution.} First, we here systematically investigate the causes of misalignment in the context of shadow models.
We do this by disentangling the impact on misalignment of the different sources of ML randomness as well as the impact of training shadow models on a different dataset.
We show that when the adversary uses a different weight initialisation for shadow models than the target model's, the former end up misaligned with the latter.
Conversely, an adversary having knowledge of the target model's initialisation is able to train shadow models which are internally much more similar to the target model.
This finding has implications for fine-tuning and model update scenarios where the base model is publicly available.
While we focus on MIAs against classification models, this fundamental finding is also relevant for any inference attack targeting DNN models, incl. attacks targeting generative models~\cite{stadler2022synthetic,houssiau2022tapas}.
Remarkably, the other sources of randomness, taken individually or combined together, and even the use of a disjoint dataset from the same distribution, do not lead to a noticeable difference between shadow and target models.

Second, we extend and evaluate the effectiveness of re-alignment methods proposed in the model fusion literature~\cite{ashmore2015method,li2015convergent} to the shadow modelling context.
Such methods aim to modify the weights of a model, without modifying its function, in order to reduce the distance between its weights and the weights of another model.
We show these methods to successfully reduce the measured misalignment between the target and shadow models.
Our results however suggest that re-alignment techniques are imperfect as they struggle with re-aligning large and middle layers.

Third, we perform a comprehensive evaluation of white-box MIAs, analysing the impact of misalignment and subsequent re-alignment techniques on the effectiveness of white-box MIAs.
We study a range of datasets, models, feature types and threat models to isolate the impact of mis/re-alignment.
We then show that re-alignment can improve the accuracy of MIAs, sometimes by a large margin, while they come at no cost to the adversary.
Indeed, they require very little compute and never perform worse than the misaligned shadow models.
Incidentally, we discover a set of features -- the input activations entering the target's label neuron -- which contain new membership signal unaccounted for by commonly used output activations and gradients~\cite{nasr2019comprehensive}.

We find that MIAs based on activations of internal layers suffer strongly from misalignment between shadow models, while MIAs based on gradients are only sometimes significantly affected.
We show that re-aligning the shadow models strongly improves the former's performance and, sometimes significantly, the latter's performance.
Hence, partially exposed models, i.e., models released without the final classification layer, e.g., for embedding purposes are more vulnerable than previously believed as the adversary cannot compute gradient features but only activation features.

We also find evidence that standard MIAs, i.e., combining all types of features and layers of misaligned shadow models, can be improved by re-alignment techniques.
On the CIFAR10 dataset with a false positive rate of 1\%, white-box MIA using re-aligned shadow models improves the true positive rate by 4.5\%.

Taken together, our results highlight that on-device deployment increase the attack surface and that the newly available information can be used by an adversary. 
The further highlight that privacy risk assessments in the white-box setting should account for the symmetries of the target model architecture when designing the attack, and that re-alignment techniques provide a simple and inexpensive means to do this while never affecting the attack performance and improving it in some settings.

%% file: sections/background.tex
\section{Problem statement}\label{sec:problem_statement}

\subsection{Deep neural networks}\label{subsec:dnn}
In this work, we mainly focus on attacks against two broad classes of deep neural networks:  multilayer perceptrons (MLP), consisting of fully connected layers, and standard convolutional neural networks\footnote{In Appendix~\ref{appendix:resnet}, we describe the ResNet~\cite{he2016deep} architecture which has a more complex structure than the standard CNN and present a re-alignment algorithm tailored to it.} (CNN), consisting of convolutional layers followed by fully connected layers.
Both can be written as a function $f$ consisting of a sequence of $L$ layers: $f=g_L \circ \ldots \circ  g_1$, with $\circ$ denoting composition.

A fully connected (FC) layer applies an affine transformation, followed by a non-linear activation function, to the input of the previous layer:
    $g_l(x^l) = \sigma_l(W^l x^{l-1} + b^l)$, with $W^l \in \mathbb{R}^{D^l \times D^{l-1}}$ a weight matrix,  
    $b^l \in \mathbb{R}^{D^l}$ a bias vector, and
    where $\sigma_l: \mathbb{R}^{D^l} \rightarrow \mathbb{R}^{D^l}$ denotes the non-linear activation function. 
Here, $x^l \in D^l$ denotes the input to the $l$-th layer (and the output of the previous layer) for every $l=1, \ldots, L$.
For the input layer, we write $x^0=x$.
Popular choices of non-linear activation functions, applied coordinate-wise, include ReLU: $x \rightarrow \max(0, x)$, $\tanh: x \rightarrow \frac{e^x-e^{-x}}{e^x+e^{-x}}$, and the sigmoid $\sigma: x \rightarrow \frac{1}{1+e^{-x}}$.
For a classification task over $N_c$ classes, the non-linear activation function used for the output layer ($l=L$) is the softmax function:
$\sigma(x^L_i) = e^{x^L_i} / \sum_{j=1}^{D^L} e^{x^L_j}$ for $i \in \{1, \ldots, D^L \}$, where $D^L$ is set equal to the number of classes $N_c$.

An FC layer $l$ consists of $D^l$ neurons, each outputting a single activation value $x^l_d, d \in \{1,\ldots,D^l\}$. 
The functionality of neuron $d$ can be described by:
(1) a vector of \textit{input weights}  $w^l_{\text{in}}(d)= (W^{l}_{d, 1}, \ldots, W^{l}_{d, D^{l-1}}, b^l_d)$ (row $d$ of the weight matrix and coefficient $d$ of the bias vector), used to compute the output activation from the outputs of the previous layer: $x^l_d = \sigma_l(b^l_d + \sum_{d'=1}^{D^{l-1}}W^l_{d,d'}x^{l-1}_{d'})$, and (2) a vector of \textit{output weights} $w^l_{\text{out}}(d)=(W^{l+1}_{1, d}, \ldots, W^{l+1}_{D^{l}, d})$, where coefficient $d'$ is used to multiply the output activation $x^l_d$ by the input weight of the $d'$-th neuron in the next layer.

In a convolutional layer, neurons are replaced with filters whose outputs are activation maps instead of single activation values.
Let $C^l$ be the number of filters in the $l$-th layer. Each filter $c \in \{ 1,\ldots, C^l \}$ is represented by a 3-D weight tensor $W^{l}[c]$ of size $C^{l-1} \times K_1^l \times K_2^l$, with $(K_1^l, K_2^l)$ the kernel size, and a bias term $b^{l}_c \in \mathbb{R}$. 
Thus, the functionality of the layer can be represented by a 4-D weight tensor $W^l$ and a 1-D vector $b^l$, after concatenating the weight tensors (and bias terms) of individual filters.
Each filter acts independently on the input $x^{l-1}$, which is 3-D (e.g., an image), by applying a convolution~\cite{goodfellow2016deep,paszke2019pytorch}:
    $x^{l}[c] = b^{l}_c + 
    \sum_{c'=1}^{C^{l-1}} W^{l}[c][c'] \ast x^{l-1}[c']
    $.
This is typically followed by non-linearity, pooling and dropout operations. 
When a convolutional layer $g_l$ is followed by an FC layer, its output $x^l$ is flattened to a 1-D vector.
In summary, the functionality of filter $d$ can be described by: (1) a vector of input weights $w^l_{\text{in}}(d)= (W^{l}[d][1], \ldots, W^{l}[d][D_{l-1}], b^l_d)$ and (2) a vector of output weights $w^l_{\text{out}}(d)=(W^{l+1}[1][d], \ldots, W^{l+1}[D_{l}][d])$, each obtained by flattening then concatenating the weight matrices.

\subsection{Threat model}\label{subsec:threat_model}
An entity (e.g., a company) trains a machine learning (ML) model to perform a classification task.
The goal of the model is to infer with high accuracy the correct label $y \in \{1, \ldots, N_c\}$ of records $x$ where $(x, y)$ is sampled from a distribution $\mathcal{D}$, with $N_c$ the number of classes.
The model $\mathcal{M}_T$ is trained on a private dataset $D_T = \{(x_1, y_1), \ldots, (x_n, y_n)\}$ of $n$ records sampled from $\mathcal{D}$ with their labels.
The model is obtained by running a training algorithm $\mathcal{T}$ on the dataset $D_T$ with randomness defined by a seed $seed_T$: $\mathcal{M}_T=\mathcal{T}(D_T, seed_T)$.\footnote{In Pytorch, $seed_T$ represents the argument passed to \texttt{torch.manual\_seed} and \texttt{torch.cuda.manual\_seed}.}
The training algorithm encompasses the choice of a network architecture $\mathcal{A}$, a loss function $\mathcal{L}$, and training hyperparameters such as learning rate and batch size.

We are interested in on-device releases, where the model is deployed on user devices, effectively as a \textit{white box} since the architecture and parameters are now accessible to the users.
We furthermore assume that each record in the dataset relates to one individual.

The key privacy question is whether a malicious user, \textit{the adversary}, can recover sensitive information about the individuals from the model. 
MIAs, which we focus on in this work, are the standard tool to assess privacy risks of ML models~\cite{shokri2017membership,jayaraman2019evaluating,jagielski2020auditing,nasr2021adversary,liu2022ml,li2022auditing}, aiming to infer whether a particular target record $(x_T, y_T)$ was used to train the \textit{target model} $\mathcal{M}_T$.
In the black-box setting, MIAs have been extensively studied, leveraging the model's confidences on $x_T$, $\mathcal{M}_T(x_T)=(\hat{y}^1, \ldots, \hat{y}^{N_c})$, or the top $k$  predictions, to infer whether $(x_T, y_T)$ is a member~\cite{shokri2017membership,carlini2022membership,ye2022enhanced}. 
In the white-box setting, the adversary additionally knows the model architecture and weights.
Given this knowledge, the adversary can design attacks \textit{at least as strong as a given black-box attack $\mathcal{B}$}, by combining the signal of the output layer (used by $\mathcal{B}$) with additional signal from the internal layers (e.g., activations or gradients).

There are mainly two threat models in the white-box setting, depending on the knowledge the adversary has about the private dataset $D_T$.

In the first threat model~\cite{nasr2019comprehensive}, the adversary knows (1) part of the training dataset $D_{\text{in}} \subset D_T$ (referred to as member records) and (2) a same-sized dataset of non-members $D_{\text{out}}$, i.e., $|D_{\text{out}}|=|D_{\text{in}}|$ and $D_{\text{out}} \cap D_T = \emptyset$.
This is a strong assumption reflecting the capabilities of an auditor.
To run the attack, first, the auditor feeds records $(x,y)\in D_{\text{in}} \cup D_{\text{out}}$ to the target model $\mathcal{M}_T$ to extract record-level features $\mathcal{F}(\mathcal{M}_T, x,y)$ such as activations of internal layers $x^l, l \in \{ 1 \ldots, L\}$ or gradients $\frac{\partial\mathcal{L}}{\partial W^l}(x, y), l \in \{ 1 \ldots, L\}$. 
Second, it labels each $\mathcal{F}(\mathcal{M}_T,x,y)$ with the corresponding membership information (member if $(x,y) \in D_{\text{in}}$ and non-member otherwise, i.e., if $(x,y) \in D_{\text{out}}$).
Third, it trains a meta-classifier (MC) to infer the label from the features.
Finally, it applies the MC to the features $\mathcal{F}(\mathcal{M}_T, x_T, y_T)$ extracted from the target model and record.

In the second, weaker threat model~\cite{leino2020stolen}, the adversary has some data, e.g., an auxiliary dataset $D_A$ from the same distribution $\mathcal{D}$, which may or may not be overlapping with the private dataset $D_T$.
The adversary does not know which of these records, if any, were used to train the model.
As such, even though the adversary can feed $(x,y) \in D_A$ to the target model $\mathcal{M}_T$ to extract features $\mathcal{F}(\mathcal{M}_T,x,y)$, it cannot label them with the correct membership information of $(x,y)$ as this information is not available.

This is where \textit{shadow models} are extremely useful, as the adversary can train one or more models $\mathcal{M}_1  = \mathcal{T}(D_1, seed_1), \ldots, \mathcal{M}_K = \mathcal{T}(D_K, seed_K)$ using the same training algorithm $\mathcal{T}$ on subsets $D_1, \ldots, D_K$ of $D_A$.
Since the adversary controls the shadow models' training data, it can build pairs of features and membership labels $\big( \mathcal{F}(\mathcal{M}_k, x, y), \mathbb{I}_{D_k}(x,y) \big)$, for every $(x,y) \in D_A, k=1,\ldots,K$, where $\mathbb{I}_S$ denotes the indicator function over dataset $S$.
The adversary can then train a meta-classifier and run the attack on the target record $(x_T, y_T)$ as before.

In this work, we focus on the second threat model as it reflects a weaker adversary, which is more realistic in the on-device setting.
We refer to it as the \textit{classical adversary}.
We make the following assumptions on its knowledge:

\underline{Assumption 1)} The adversary knows the training algorithm $\mathcal{T}$, allowing it to train a similar model from scratch on datasets of its choice.
    We make this assumption to simplify the analysis, similarly to prior works~\cite{ateniese2015hacking,nasr2019comprehensive,leino2020stolen,zhang2021leakage,suri2022formalizing}.
    While the model architecture is revealed to the users (on device) and the loss function is the standard cross-entropy in most cases, the hyperparameters may not always be available to the adversary.
    They would be, e.g., when the code to train the model or a full description of $\mathcal{T}$ is available.

\underline{Assumption 2)} The adversary \textit{does not have access to the same randomness (seed)} as the one used to train the target model, i.e., the adversary does not know $seed_T$.
Thus, we will assume the shadow models to be trained using a different seed $seed_k \neq seed_T, k=1,\ldots, K$.
Indeed, training scripts are often not seeded at all, or they are seeded but the target seed $seed_T$ is not hard-coded nor made available.
In the rare cases where $seed_T$ is published along with the code, ensuring reproducibility in neural network libraries is challenging due to non-determinism in the hardware~\cite{chen2022towards}.
Assuming that the adversary generates $seed_k, k=1,\ldots,K$ randomly, the chances of collision with $seed_T$ are negligible.

Under these assumptions, shadow models differ from the target model in the following factors:
\begin{enumerate}
\item \underline{Training dataset.} 
    The training dataset of shadow models can be partially overlapping with $D_T$~\cite{carlini2022membership} or disjoint from $D_T$~\cite{shokri2017membership,ganju2018property}. 

\item \underline{Weight initialisation.} This is the value of the model weights at the beginning of training, typically sampled randomly and independently from a distribution centred around zero, e.g., uniform or normal with parameters dependent on the layer sizes~\cite{glorot2010understanding,paszke2019pytorch}.

\item \underline{Batch ordering}. This is the ordering of mini-batches on which the gradient is estimated in each training step.
    At the beginning of each epoch (representing a complete pass over the dataset), the training dataset is randomly shuffled.
    Thus, the use of different training randomness between the shadow models and the target model leads to different batch ordering.

\item \underline{Dropout selection.} Dropout~\cite{srivastava2014dropout} is a popular  regularisation technique aiming to reduce overfitting. 
    Dropout is applied to the outputs of the internal layers.
    In each forward pass, dropout randomly sets each value of an input vector independently to 0 with probability $p$, or multiplies it by $\frac{1}{1-p}$ with probability $1-p$.
    The use of different randomness to train shadow models and the target model leads to different values being ``zeroed out''. 
\end{enumerate}

While this list is exhaustive for the neural networks considered in our analysis of the causes of shadow model misalignment (Sec.~\ref{sec:identifying_causes}), additional sources of randomness could be considered.
For instance, the DP-SGD algorithm, commonly used to train models with formal privacy guarantees~\cite{dwork2006calibrating}, randomly perturbs gradients during training~\cite{abadi2016deep}.

The standard way to control the randomness of factors (2)-(4) is to use a common pseudorandom number generator (PRNG), setting its seed at the beginning of training.
In Sec.~\ref{sec:identifying_causes}, we will measure the influence of each factor by using a separate PRNG for each factor.

\subsection{Symmetries in deep neural networks}\label{subsec:symmetries}
Deep neural networks are known to exhibit symmetries~\cite{hecht1990algebraic,ashmore2015method,li2015convergent,yurochkin2019bayesian,tatro2020optimizing,singh2020model,wang2020federated,entezari2021role}, which are essential for understanding the challenges of white-box attacks.
Consider a deep neural network $f_{\theta}$ implementing an architecture $\mathcal{A}$.
Here, $\theta = ((W^1, b^1), \ldots, (W^L, b^L)) \in \mathcal{W}$ denotes its weight representation, with $\mathcal{W}$ the space of values that can be taken by the weights.
\begin{definition}[Symmetry]\label{def:symmetry}
Let $\mathcal{A}$ be an architecture with weight space $\mathcal{W}$. 
A symmetry is a mapping $T: \mathcal{W} \rightarrow \mathcal{W}$ that for every $\theta \in \mathcal{W}$ preserves the function $f_{\theta}$: $ \forall \theta \in \mathcal{W}, \forall x, f_{T(\theta)}(x)=f_{\theta}(x)$. 
\end{definition}
We denote by $\mathcal{S}(\mathcal{A},\mathcal{W})$ the set of symmetries over architecture $\mathcal{A}$ and weight space $\mathcal{W}$.
The main types of symmetries are:
\begin{enumerate}
\item \underline{Permutation equivalence:} The ordering of the neurons (filters) of internal layers can be arbitrarily permuted without changing $f$, as long as their corresponding input weights, bias, and output weights are permuted accordingly.
For instance, in an FC layer $g_l$, applying a permutation to the rows of $W^l$, the bias vector $b^l$, and the columns of $W^{l-1}$ does not change $f$. 

\item \underline{Scale invariance:} The input weights and the bias of a neuron can be multiplied by a constant $c>0$ without changing $f$, so long as the output weights are multiplied by $1/c$. 
This holds when there is no non-linearity, or when the non-linearity is the ReLU function.

\item \underline{Antisymmetry:} 
The input weights, bias, and output weights of a neuron can be multiplied by -1 without changing $f$ if the non-linearity is antisymmetric, e.g., $\forall x, \tanh(x)=-\tanh(-x)$.
\end{enumerate}

\textbf{Detailed examples:} Consider an architecture of two linear layers, where the first layer is followed by the ReLU non-linearity: $\theta=( (W^1, b^1), (W^2, b^2) )$, with $W^1 \in\mathbb{R}^{3\times2}$, $W^2 \in\mathbb{R}^{4\times3}$, $b^1\in\mathbb{R}^{3}$ and $b^2\in\mathbb{R}^4$, i.e., $\mathcal{W}=\mathbb{R}^{3\times2}\times\mathbb{R}^{3}\times\mathbb{R}^{4\times3}\times\mathbb{R}^4$.
The operation of the first layer is $x^1_i = \max(0, W^1_{i1}x^0_1 + W^1_{i2}x^0_2 + b^1_i), i \in \{ 1, 2, 3\}$.
The operation of the second layer is $x^2_j = W^2_{j1}x^1_j + W^2_{j2}x^1_2 + W^2_{j3}x^1_3 +b^2_j, j \in \{ 1, 2, 3, 4\}$.
This architecture has permutation equivalence symmetries because for any permutation $\sigma$ over $\{1, 2, 3\}$, the network $\theta'= ( (W'^1, b'^1), (W'^2, b'^2) )$, with $W'^1= (W^1_{\sigma(i)j})_{1\leq i \leq 3,1\leq j \leq 2}$, $b'^1=(b_{\sigma(i)})_{i=1,\ldots,3}$, $W'^2=(W^2_{k\sigma(i)})_{1\leq k \leq 4, 1\leq i \leq 3}$ and $b'^2=b^2$ satisfies $f_{\theta'}(x) = f_{\theta}(x), \forall x$ (see ~\cite{ganju2018property} for a proof).
It also has scale invariance symmetries: if we replace $W^1$ with $W'^1= (c_i W^1_{ij})_{1\leq i \leq 3, 1\leq j \leq 2}$, $b'^1 $ with $(c_i b_i)_{1 \leq i \leq 3}$ and $W^2$ with $W'^2=(\frac{1}{c_i} W^2_{k\sigma(i)})_{1\leq k \leq 4, 1\leq i \leq 3}$, where $c_i>0, i=1,\ldots,3$, then $f_{\theta'}(x) = f_{\theta}(x), \forall x$.
If the non-linearity is tanh instead of ReLU, the architecture has antisymmetry instead of scale invariance symmetries, i.e., if we replace $W^1$ with $-W^1$, $b^1$ with $-b^1$, and $W^2$ with $-W^2$, then $f_{\theta'}(x) = f_{\theta}(x), \forall x$.

\subsection{Misalignment in shadow models}\label{subsec:misalignment}
The existence of symmetries, coupled with differences in how shadow models are trained respectively to the target model, leads to shadow models which are misaligned respectively to the target model.
We define this concept below:

\begin{definition}[$d$-misalignment]
    Let $\mathcal{A}$ be an architecture with weight space $\mathcal{W}$. 
    Let $f_{\theta_1}$ and $f_{\theta_2}$ be two models of architecture $\mathcal{A}$ with corresponding weights $\theta_1, \theta_2\in\mathcal{W}$.
    We assume the models to be trained for solving the same task but under different conditions, such as using different datasets, randomness or training algorithms.
    Let $d$ be a distance metric over $\mathcal{W}$.
    We say that $\theta_1$ is $d$-misaligned with respect to $\theta_2$ if there exists a symmetry $T$ that reduces its weight distance with respect to $\theta_2$, i.e., $\exists T\in \mathcal{S}(\mathcal{A},\mathcal{W}): d(T(\theta_1), \theta_2) < d(\theta_1, \theta_2)$.
\end{definition}

\begin{definition}[$d$-alignment]
    We say that $\theta_1$ is $d$-aligned respectively to $\theta_2$ if for all symmetries $T$, it is not possible to reduce the weight distance with respect to $\theta_2$, i.e., $\forall T\in \mathcal{S}(\mathcal{A},\mathcal{W}): d(T(\theta_1), \theta_2) \geq d(\theta_1, \theta_2)$.
\end{definition}

The notion of ``(mis)alignment'' is only meaningful if the two models learn similar features.
This is why our definition requires the two models $f_{\theta_1}$ and $f_{\theta_2}$ to be trained for solving the same task.
Indeed, one could imagine a model trained to classify animal species and another model trained to classify vehicles; since the models likely learn different features, the notion of ``(mis)alignment'' is ill-defined.

In our experiments, we use the Euclidean distance between weights as the distance metric. For simplicity, we refer to models as ``(mis)aligned'' instead of ``$d$-(mis)aligned''.

While it may be theoretically possible to train shadow models which are aligned with the target model, in practice shadow models always end up misaligned. We illustrate this phenomenon by training a standard CNN ~\cite{leino2020stolen}, which consists of two convolutional layers followed by two fully connected layers (see Sec.~\ref{subsec:misalignment_experimental_setup} for the complete details).

Fig.~\ref{fig:visualise_misalignment}b and c show the internal representations of a target model $\mathcal{M}_T$ as well as those of shadow models trained by the classical adversary.
This adversary has no knowledge of the seed used to train the target model, and uses a different seed for each shadow model.
To simulate this setting, we randomly sample four mutually disjoint subsets of the CIFAR10 dataset~\cite{krizhevsky2009learning} of size 12500, using the first subset to train a target model $\mathcal{M}_T$ and the others to train three shadow models.
We visualise the activations maps, computed on an image of an aeroplane, of the first convolutional layer of each model.
The activation maps describe the functionality of the convolutional layer.
Each activation map is a matrix of real values that are non-negative, due to the ReLU non-linearity function applied by the convolutional layer.
We scale the values between 0 and 1 independently in each map, depicting 0 as black and 1 as white.
The values indicate which parts of the image an activation map focuses on and which parts are ``zeroed out'' by the computation.

Fig.~\ref{fig:visualise_misalignment}b and c show that while the target and shadow models learn many similar features (examples are highlighted in colour),  these are not located in the same position consistently across models.
This phenomenon also affects the upper, fully connected layers, but these cannot be visualised because their activations are scalars.

Importantly, \textit{misalignment does not affect the output classification layer}.
Even though the outputs of shadow models might differ, they have the same functionality for target and shadow models: the $l$-th output consistently encodes the probability that the input image belongs to the $l$-th class.
This explains why black-box features extracted from shadow models perform very well, as reported in many works~\cite{shokri2017membership,choquette2021label,zhang2021leakage,carlini2022membership}.

As for white-box features, misalignment reduces the ability of the meta-classifier to effectively perform the inference task. 
We demonstrate this by training white-box MIAs against the CNN architecture described above using \textit{features extracted from an internal layer}.
More specifically, we use as features the activations of the second to last layer $x^{L-1}$.
To understand whether these features have signal useful for the MIA task, we train an attack on features extracted from the target model (implicitly assuming the auditor threat model of Sec.~\ref{subsec:threat_model}).
This attack reaches a high AUC of 0.805, meaning that the features hold signal.
We then train an attack on features extracted from $K=10$ shadow models. 
This attack achieves an Area Under the Curve (AUC) of 0.509, comparable to a random guess.
The two experiments illustrate that the second to last layer contains signal useful for the membership task, but that this signal is not directly accessible from the shadow models through a na\"ive application of the shadow modelling technique.

Finally, Fig.~\ref{fig:visualise_misalignment}d shows the features of shadow models after applying one of the re-alignment techniques evaluated in this work, correlation-based matching~\cite{li2015convergent} (described in Sec.~\ref{sec:aligning_layers}).
It is clear that re-alignment leads to more similar features across the same position in the shadow models resp. to the target model.

%% file: sections/related_work.tex
\section{Related work}\label{sec:related_works}
\textbf{Misalignment} has been studied in the model fusion literature where there is a need to train multiple models and then combine the models into a single one. 
Model fusion has e.g. been used to achieve the benefits of ensembling (better accuracy) without having to store multiple models~\cite{ashmore2015method,singh2020model}, to enable crossover mutations over neural networks in evolutionary algorithms~\cite{ashmore2015method}, and to average models in federated learning~\cite{yurochkin2019bayesian,wang2020federated}.
Misalignment has also been used to better understand neural network representations~\cite{li2015convergent} and the loss landscapes of stochastic gradient descent~\cite{tatro2020optimizing,entezari2021role}.
Re-alignment techniques have been proposed, e.g.,~\cite{ashmore2015method,li2015convergent}, which we extend and apply for the first time to shadow models.

A handful of works have identified misalignment to be an issue for property inference attacks (PIA), a specific use of shadow models, and proposed solutions. 
Ganju et al.~\cite{ganju2018property} looked at  permutation equivalence-type symmetries in MLP models, and showed them to affect the performance of the naïve PIA meta-classifier based on the model weights.
The authors propose two approaches to tackle this challenge: weight-based neuron sorting and set-based meta-classifiers.
Recently, Suri and Evans extended set-based meta-classifiers to CNNs~\cite{suri2022formalizing}.
Differently from these works, we systematically analyse the root causes of misalignment in shadow models and explore the effectiveness of re-alignment techniques. 
We then analyse the impact of misalignment and re-alignment techniques on white-box MIAs, the standard privacy test. 
More specifically, we show weight-based neuron sorting to be ineffective in reducing the misalignment and improving MIAs, and extend set-based classifiers to MIAs, showing them to perform worse or on par with re-alignment.

\textbf{Impact of weight initialisation.} Balle et al.~\cite{balle2022reconstructing}. developed an attack using shadow models to reconstruct a training record based on the model weights. 
They found that their attack fails if the shadow models are not trained using the same weight initialisation as the target model.
Similarly, Jagielski et al.~\cite{jagielski2020auditing} obtained tighter estimates of the privacy of the DP-SGD algorithm~\cite{abadi2016deep} when using a fixed weight initialisation.
Our work provides an explanation for these empirical findings through the misalignment perspective and suggests re-alignment as a potential solution to improve attacks when the adversary does not know the weight initialisation.

\textbf{Membership inference attacks} (MIA) have been studied extensively in the black-box setting, see e.g.,~\cite{shokri2017membership,salem2018ml,choquette2021label,truex2019demystifying,carlini2022membership,li2022auditing}.
The features typically used are the model confidences on a record $x$ with the main vulnerability exploited being that the model tends to be more confident on training records compared to unseen records.
Together, these works have built a solid understanding of the conditions under which black-box MIAs are most likely to succeed, such as when the model is overfitted~\cite{shokri2017membership,yeom2018privacy}, when the targeted records are outliers or members of subpopulations~\cite{long2020pragmatic,kulynych2022disparate}, and when the auxiliary data comes from the same distribution as the training dataset and is more abundant~\cite{salem2018ml,truex2019demystifying}.

Comparatively very few works have developed white-box MIAs.
Nasr et al.~\cite{nasr2019comprehensive} studied the auditor setting (described in Sec.~\ref{sec:related_works}), 
training a meta-classifier (MC) on activations and gradients extracted \textit{directly from the target model}.
Shadow models are, however, not used in this setting, since the signal available through the target model is, by definition, perfectly aligned.

Leino and Fredrikson~\cite{leino2018influence} considered an adversary who has access to a disjoint set of records $D_A\sim\mathcal{D}$, with $D_A \cap D_T=\emptyset$.
The authors identify the issue of misalignment and address it by training \textit{proxy} models that share the same internal representation as $\mathcal{M}_T$ in all the layers up to and including a given layer $l$. Their method is, however, computationally heavy. 
The proxy models are first: (1) initialised with the target model's weights up to layer $l$ and frozen during training and (2) initialised randomly in the remaining, upper layers.
The function applied by the upper layers to each record is then linearly approximated using influence functions~\cite{leino2018influence}.
Finally, a meta-classifier is trained on a measure of behaviour difference between the linear approximations of proxy and target models. 
The adversary needs to repeat this procedure by training shadow models to mimic the role of the target model and by training proxy models for each shadow model and layer.
For computational reasons, we here instantiate their approach with the last layer and find that it performs worse than the na\"ive attack using the same features.

Sablayrolles et al.~\cite{sablayrolles2019white} aim to answer the question of whether the membership signal of internal layers is redundant with the model confidences. If they were completely redundant, white-box attacks could indeed never outperform black-box attacks. 
Under specific assumptions on the distribution of the model weights, they derive that the optimal attack only depends on the loss function.
This implies that white-box attacks could not exploit additional membership signal to the optimal black-box attack, since the loss can be estimated from black-box outputs.
These results however contradict empirical work that showed that white-box attacks \textit{can} perform better than black-box, reporting double-digit accuracy gaps~\cite{li2022auditing}. 
Further work is thus necessary to understand when the assumptions made by Sablayrolles et al. hold in practice.

%% file: sections/causes_misalignment.tex
\section{Causes of shadow model misalignment}\label{sec:identifying_causes}
We now analyse the causes of misalignment in shadow models trained by the adversary who does not know the randomness used to train the target model $\mathcal{M}_T$.

Shadow models typically differ from the target model in the training dataset, weight initialisation (WI), batch ordering (BO) and dropout selection (DS).
To understand the impact of these individual factors, we use a separate PRNG to control their randomness $seed_T=(seed_T^{WI}, seed_T^{BO}, seed_T^{DS})$, changing one seed a time between the target model and shadow models.
For instance, to measure the impact of weight initialisation, we train $K$ shadow models $\mathcal{M}_k=\mathcal{T}(D_T, seed_k^{WI}, seed_T^{BO}, seed_T^{DS})$, with $seed^{WI}_k \neq seed_T, k \in \{1,\ldots,K \}$
We then measure the resulting misalignment in each layer using the metrics described in Sec.~\ref{subsec:misalignment_metrics}.
Then, we change multiple factors together, progressively increasing the adversary uncertainty, up to our adversary of interest for which all selected factors are different between shadow and target models: $D_k\neq D_T, seed_k^{WI}\neq seed_T^{WI}, seed_k^{BO}\neq seed_T^{BO}, seed_k^{DS}\neq seed_T^{DS}$.
From a security perspective, this means that we explore intermediate attack scenarios which may lead to lower levels of misalignment than  the one measured for our adversary (Sec.~\ref{subsec:misalignment_experimental_setup} and ~\ref{subsec:causes_results}).

\subsection{Metrics to measure the misalignment}\label{subsec:misalignment_metrics}
To measure the misalignment between a shadow model $\mathcal{M}$ and the target model $\mathcal{M}_T$, we compute in each layer $l$ the weight misalignment score (WMS). The WMS is defined as the Euclidean distance between the weights and biases of neurons (filters) of $\mathcal{M}$ and $\mathcal{M}_T$ in layer $l$: 
$\sqrt{|| W^l(\mathcal{M}_T) - W^l(\mathcal{M}) ||_2^2 + || b^l(\mathcal{M}_T) - b^l(\mathcal{M})||_2^2 }$.
Here, we denote by $W^l(\mathcal{M})$ the weight matrix of the $l$-th layer of $\mathcal{M}$ and by $b^l(\mathcal{M})$ its bias vector. In a convolutional layer, the metric is defined analogously over the weights and bias of individual filters.
We select this metric for its simplicity, interpretability, and to cover the functionality of a neuron (filter) in terms of weights.
In Appendix~\ref{appendix:misalignment_metric}, we present and discuss results obtained with two additional metrics based on the output activations.

\textbf{Random permutation.} 
To contextualise the change incurred by each factor in a layer, we compute as a baseline the WMS between the target model and the same model where a random permutation is applied to the neurons (filters) of the layer.

\begin{table*}[!htbp]
\centering
\caption{Weight misalignment scores. 
Weight initialisation (WI) is the most influential factor responsible for misalignment. 
We highlight in bold the classical adversary (All $\neq$) and the one knowing the target model initialisation ($\neq$ BO, $\neq$ DS, and DD).
}
\begin{tabular}{lcccc}
\toprule
\multirow{2}{*}{\textbf{Difference w.r.t. target model $\mathcal{M}_T$}}  & \textbf{First layer} & \textbf{Second layer} & \textbf{Third layer} & \textbf{Fourth layer}\\
   & \textbf{(20 filters)} & \textbf{(50 filters)} & \textbf{(500 neurons)} & \textbf{(10 neurons)} \\
\midrule
\multicolumn{5}{c}{\textbf{Weight misalignment scores ($\downarrow$ indicates better alignment)}} \\
\midrule
Random permutation & 11.81 (0.51) & 16.58 (0.15) & 31.87 (0.06) & 12.80 (0.66) \\
\midrule
$\neq$ Weight initialisation (WI) & 12.09 (0.34) & 16.24 (0.20) & 30.46 (0.45) & 12.54 (0.14) \\
$\neq$ Batch ordering (BO) & 3.42 (0.34) & 8.88 (0.26) & 21.51 (0.85) & 6.62 (0.15) \\
$\neq$ Dropout selection (DS) & 3.52 (0.30) & 8.72 (0.23) & 21.25 (0.46) & 6.47 (0.16) \\
\midrule
Overlapping datasets & 4.32 (0.36) & 10.52 (0.31) & 23.20 (0.60) & 7.23 (0.11) \\
Disjoint datasets (DD) & 5.11 (0.59) & 11.21 (0.11) & 24.24 (0.35) & 7.56 (0.05) \\
\midrule
\textbf{$\neq$ BO, $\neq$ DS, and DD} & 5.17 (0.31) & 11.46 (0.03) & 24.92 (0.21) & 7.55 (0.07) \\
$\neq$ MI, $\neq$ BO, and $\neq$ DS & 12.05 (0.43) & 16.30 (0.21) & 30.53 (0.34) & 12.65 (0.18) \\
\textbf{All $\neq$ ($\neq$ WI, $\neq$ BO, $\neq$ DS, and DD)} & 12.44 (0.36) & 16.56 (0.20) & 30.98 (0.50) & 12.87 (0.28) \\
\bottomrule
\end{tabular}
\label{table:misalignment_scores_causes}
\end{table*}

\subsection{Experimental setup}\label{subsec:misalignment_experimental_setup}
\textbf{Dataset.} We perform experiments on the CIFAR10 dataset~\cite{krizhevsky2009learning}, which consists of 60000 $32\times 32 \times 3$ colour images labeled with one of 10 classes.
The dataset was originally partitioned into a train and test splits of 50000 and 10000 images, respectively.
We randomly sample 12500 records from the train split as our target dataset $D_T$ and three mutually disjoint subsets (five random subsets) of same size for experiments using shadow models trained on disjoint datasets (overlapping datasets).
In the case of overlapping datasets, the overlap between two different models is of roughly 25\%.

\textbf{Model architecture and training details.} We use the CNN architecture of Leino and Fredrikson~\cite{leino2018influence}, consisting of two convolutional layers and two fully connected layers.
For better utility and to limit overfitting, differently from this work we use a smaller dropout probability at the output of the fully connected layer ($p=20\%$), a smaller learning rate, and early stopping.
We train the network using the Adam optimiser~\cite{kingma2014adam} a batch size of 64, a learning rate of $\eta=0.01$, dividing it by 2 after $E=5$ epochs of non-increasing accuracy on a validation set of 5000 random records from the test split.
We stop the training when $\eta$ drops below $10^{-5}$ or after 100 epochs, whichever comes first.

\subsection{Results}\label{subsec:causes_results}

Table ~\ref{table:misalignment_scores_causes} shows that \textbf{the main cause for the misalignment of shadow models trained by the adversary is their different weight initialisation}.
When shadow models are trained on the same dataset, with the same batch ordering and dropout selection as the target model, but with a \textit{different weight initialisation}, the weight misalignment scores reach values comparable to applying a random permutation (RP) to the neurons (filters) of the target model $\mathcal{M}_T$.
In the first layer, the score computed over shadow models having a different weight initialisation ($\neq$ WI) is 12.09, similar to applying a random permutation to the filters of the target model (11.81).
In contrast, shadow models differing only in their batch ordering ($\neq$ BO) or dropout selection ($\neq$ DS) are much less misaligned to the target model, reaching scores of 3.42 and 3.52, respectively.

Interestingly, shadow models differing from the target model in their training dataset (but not in the other factors) are also broadly aligned to the target model, reaching a weight misalignment score of 4.32 and 5.11 when trained on a dataset that can be overlapping with, respectively is disjoint from, $D_T$.
Fig.~\ref{fig:changing_seed} and ~\ref{fig:changing_dataset} in Appendix~\ref{appendix:visualisation} illustrate that the activation maps of the first convolutional layer are indeed visually aligned to those of the target model when changing only the batch ordering, dropout selection, or the training dataset.

We also instantiate concrete attack scenarios by changing multiple factors together.
More specifically, Table ~\ref{table:misalignment_scores_causes} shows that changing everything but the WI ($\neq$ BO, $\neq$ DS, and disjoint dataset) results in a score of 5.17, much lower than our adversary of interest which additionally uses a different WI (12.44). 
\textbf{From a security perspective, this strongly suggests that adversaries having knowledge of the weight initialisation can train shadow models which, internally, are much more similar to the target model than adversaries not having this knowledge.}
This assumption holds whenever the seed is available to the adversary, or when the target model is fine-tuned and its pretrained weights are available, or when the adversary accesses a target model before a mini-batch update is performed~\cite{salem2020updates,jagielski2022combine}.
Fig.~\ref{fig:changing_all_but_model_seed} shows that the activation maps of these shadow models are overall well aligned to the target model.
However, since now multiple factors are different, we also observe some differences, notably in columns 1 and 7.

The trends identified in the first layer also hold for the other layers.
Yet, we observe smaller gaps between $\neq$ WI compared to e.g., $\neq$ BO.
The ratio between the WMS decreases from $12.09/3.42\approx 3.4\times$ in the first layer to $1.9\times$ and $1.4\times$  in the second, resp. third layer.
This suggests that in some of the layers, even the smallest change can result in large weight differences compared to the target model.

We observe similar trends using different metrics for measuring misalignment (Table~\ref{table:misalignment_scores_causes_other_metrics}), different training hyperparameters (Tables~\ref{table:wms_comparison_dataset_sizes}-~\ref{table:wms_comparison_hyperparameters}), as well as different datasets and models (Tables~\ref{table:cifar100:misalignment_scores_causes}-\ref{table:purchase100:misalignment_scores_causes}).
We refer the reader to  Appendix~\ref{appendix:misalignment_metric} for a discussion of these results.
A particularly salient finding is that shadow models always learn a different classifier compared to the target model. 
Table~\ref{table:misalignment_scores_causes_other_metrics}A indeed shows that even though the output activations of the last layer are aligned (the $d$-th output neurons encodes the output probability for the $d$-th class), the activation misalignment scores are much larger than 0, ranging between 0.27 ($\neq$ DS) and 0.42 (classical adversary).

We also find evidence that feature misalignment is not impacted by the use of training datasets from different distributions.
Consider an adversary who knows the target model's task but not the precise training distribution.
We aim to understand whether shadow models are more misaligned when trained on a different distribution, and if they learn different features from the target model.
We train a target model on 100\% older faces of the CelebA dataset~\cite{liu2015faceattributes} to infer whether a person is smiling.
We then train shadow models either on the same distribution or on a different distribution (only 22.6\% older faces) (see Appendix~\ref{appendix:celeba} for details).
Table~\ref{table:celeba} shows similar WMS (and hence, levels of misalignment) between the two scenarios, while Fig.~\ref{figure:celeba} suggests that the features learnt are similar.

%% file: sections/alignment.tex
\section{Re-aligning the layers}
\label{sec:aligning_layers}
Our findings suggest that permutation equivalence is the main type of symmetry responsible for the misalignment between shadow models trained by the classical adversary.
Indeed, the misalignment measured in these models is comparable to applying a random permutation to internal layers of $\mathcal{M}_T$. 
A natural question is whether the internal layers of shadow models can be re-aligned to those of the target model, by seeking the ``right'' way to permute the neurons (filters) so as to minimise misalignment.
While there may not be a one-to-one mapping between the neurons, i.e., some neurons in a shadow model may be functionally different from all the neurons in the target model, correctly re-aligning even a subset of neurons may significantly reduce misalignment and improve MIA performance.

We here evaluate several re-alignment techniques from the ML literature~\cite{ashmore2015method,li2015convergent} 
that are instantiations of the bipartite assignment problem.
More specifically, we keep the target model $\mathcal{M}_T$ fixed and search for a symmetry $T$ consisting of a sequence of  permutations, $\pi_1, \ldots, \pi_L$, one for each layer, that will be applied to neurons 
(filters) in the $l$-th layer of the shadow model $\mathcal{M}$  so as to maximise its alignment to the $l$-th layer of $\mathcal{M}_T$.
The permutations can be applied in a bottom-up or top-down order.

\textbf{Computing the optimal permutation in layer $l$.}
Let $f_1, \ldots, f_{D_l}$ be the features representing the neurons (filters) $n_1, \ldots, n_{D_l}$ of layer $l$ in $\mathcal{M}$ and let $f_1^T, \ldots, f_{D_l}^T$ be the features representing the neurons (filters) $n_1^T, \ldots, n_{D_l}^T$ of layer $l$ in $\mathcal{M}_T$.
We seek the optimal permutation of neurons of $\mathcal{M}$, $\pi_l(n_1), \ldots, \pi_l(n_{D_l})$, i.e., maximising the alignment between neurons: $
\pi_l = \argmax_\pi \sum_{1\leq d \leq D_l} sim(\pi(n_d), n_d^T)$,
where $sim$ is a similarity score based on the features representing the neurons.
We explore three options for computing $sim$:

a) $sim(n_i, n_j) = - || f_i - f_j^T ||_2$, with $f_i=w_{\text{in}}^l(i)$, i.e., the input weights of neuron $i$ in $\mathcal{M}$ or $f_i=w_{\text{out}}^l(i)$, i.e., the output weights of neuron $i$ in $\mathcal{M}$.
We pick one or the other depending on whether re-alignment is performed in a bottom-up or top-down fashion, as detailed in the next paragraph.
$f^T_j$ is analogously defined except that it is computed on $\mathcal{M}_T$.

b) $sim(n_i, n_j) = - || f_i - f_j^T ||_2$, where $f_i=\big( (x_1^l)_i, \ldots, (x_R^l)_i\big)$, i.e., the output activations of neuron $i$ in $\mathcal{M}$ over records $x_1, \ldots, x_R$.
$f_j^T$ is analogously defined and computed using the same records except that it is computed on $\mathcal{M}_T$. 

c) The Pearson correlation coefficient between the output activations of the neurons, computed from $R$ records~\cite{li2015convergent}.
We refer the reader to Appendix~\ref{appendix:alignment_other_metrics} for a complete description.

We use the Hungarian algorithm~\cite{kuhn1955hungarian} to find optimal permuta- tion $\pi_l$ and $R=500$ records to compute similarity scores b) and c).

\begin{table*}[!ht]
\centering
\caption{Weight misalignment scores after applying re-alignment techniques to shadow models trained by the classical adversary.  
}
\begin{tabular}{lcccc}
\toprule
\multirow{2}{*}{\textbf{Re-alignment technique}}  & \textbf{First layer} & \textbf{Second layer} & \textbf{Third layer} & \textbf{Fourth layer}\\
 & \textbf{(20 filters)} & \textbf{(50 filters)} & \textbf{(500 neurons)} & \textbf{(10 neurons)} \\
\midrule
\multicolumn{5}{c}{\textbf{Weight misalignment scores ($\downarrow$ indicates better alignment)}} \\
\midrule
(A0) No re-alignment & 12.44 (0.36) & 16.56 (0.20) & 30.98 (0.50) & 12.87 (0.28) \\
\midrule
(A1) Weight sorting~\cite{ganju2018property}  & 11.51 (0.48) & 16.28 (0.29) & 30.96 (0.50)  & 12.74 (0.24) \\
(A2) Re-alignment after initialisation & 12.49 (0.16) & 16.00 (0.14) & 30.39 (0.53) & 8.56 (0.21) \\
(A3) Bottom-up weight-based re-alignment  & {\bf 8.87} (0.56) & {\bf 14.19} (0.19) & {\bf 29.35} (0.44) & 11.01 (0.21) \\
(A4) Top-down weight-based re-alignment & 10.80 (0.81) & 14.98 (0.24)  & 30.37 (0.48) & {\bf 5.71} (0.04) \\
(A5) Activation-based re-alignment  & 9.45 (1.00) & 14.51 (0.07) & 30.13 (0.45) & 10.54 (0.01) \\
(A6) Correlation-based re-alignment~\cite{li2015convergent} & 9.38 (0.95) & 14.58 (0.11) & 30.45 (0.49) & 11.49 (0.06) \\
\bottomrule
\end{tabular}
\label{table:misalignment_scores_alignment}
\end{table*}

\textbf{Bottom-up re-alignment.} For each layer $l=1, \ldots, L-1$, we perform the following:
\begin{itemize}
    \item[(B1)] We compute the optimal permutation $\pi_l$ that should be applied to neurons in the $l$-th layer of $\mathcal{M}$ to maximise alignment with $\mathcal{M}_T$ as per $sim$. 
    Note that under a), we use as features the \textit{input weights}, since the \textit{output weights} point towards neurons in layer $l+1$ that are not yet re-aligned.
    \item[(B2)] In the current layer $l$, we permute the rows of the weight matrix $W^l(\mathcal{M})$ and bias vector $b^l(\mathcal{M})$ according to $\pi_l$.
    \item[(B3)] In the next (upper) layer $l+1$, we permute the columns of the weight matrix $W^{l+1}(\mathcal{M})$ according to $\pi_l$.
\end{itemize}

\textbf{Top-down re-alignment.} For each layer $l=L, \ldots, 2$, we perform the following:
\begin{itemize}
    \item[(T1)] Same as (B1), except that under a), we use as features the \textit{output weights} of the neuron, since the \textit{input weights} are not yet re-aligned between $\mathcal{M}$ and $\mathcal{M}_T$.
    \item[(T2)] In the current layer $l$, we permute the columns of the weight matrix $W^l(\mathcal{M})$.
    \item[(T3)] In the next (lower) layer $l-1$, we permute the rows of the weight matrix $W^{l-1}(\mathcal{M})$ and bias vector $b^{l-1}(\mathcal{M})$.
\end{itemize}

Note that bottom-up weight-based re-alignment leads to different internal representations of $\mathcal{M}$ compared to top-down weight-based re-alignment, while bottom-up activation-based (correlation-based) re-alignment leads to the same result as top-down activation-based (correlation-based) re-alignment.
Indeed, the similarity scores under b) and c) only depend on the output activations of the $l$-th layer, which are not impacted by the permutation applied to the previous layer (i.e., $l-1$ for bottom-up order and $l+1$ for the top-down order).
However, the layer order impacts the final result under a), as the ordering of weights of a neuron is impacted by the permutation applied to the previous layer. 

These algorithms can be easily extended to convolutional layers, replacing neurons by filters and activations by activation maps.
However, special care is required when propagating a permutation to the next layer if the two layers are of different types, i.e., at the junction between convolutional and FC layers in CNNs.
To see how this impacts the algorithms, take for instance bottom-up alignment.
If the current layer (to be permuted) $l$  is convolutional while the upper layer $l+1$ is fully connected, steps (B1) and (B2) would be the same, except that they would be applied to filters instead of neurons.
However, we modify step (B3) as follows.
First, we group the columns of the weight matrix $W^{l+1}(\mathcal{M})$ by the input filter, reversing the flattening operation applied to the activation maps before they are fed to the FC layer.
Second, we permute the groups of columns  according to $\pi_l$.
Finally, we flatten the groups back to a list of columns, thus restoring the 2-D shape of the matrix.

We also present in Appendix~\ref{appendix:resnet} an extension of top-down re-alignment for ResNet architectures which are structured differently from the standard CNNs described so far, due to their use of \textit{skip connections}~\cite{he2016deep}, thus requiring a different treatment.

Note that \textit{perfect re-alignment is not guaranteed}, due to (1) the heuristic metrics used and (2) potential errors being propagated, and accumulated, from the first to the last layer being re-aligned.

\textbf{Re-alignment after initialisation.} Guided by the insight that weight initialisation is the main factor responsible for shadow model misalignment, we also experiment with shadow model re-alignment after their initialisation.
We use top-down weight-based re-alignment due to its better overall performance (see Sec.~\ref{subsec:results_mia}).

\textbf{Baseline.} We further implement the weight sorting technique of Ganju et al.~\cite{ganju2018property}. 
This approach maps a model to its \textit{canonical representation}, by sorting its layers in a bottom-up order, according to the sum of input weights of neurons.
We extend this approach to convolutional layers, and map the shadow models as well as the target model to their canonical representation.

Table~\ref{table:misalignment_scores_alignment} shows that one-to-one matching techniques (A3)-(A6) can effectively reduce the misalignment consistently.
As expected, top-down weight-based re-alignment (A4) performs best in the upper layers, perhaps due to cumulative error effects as the algorithm progresses to the lower layers (with similar trends for bottom-up re-alignment).
In particular, it achieves the most spectacular improvement relatively to the no re-alignment baseline, reducing the weight misalignment score from 12.87 to 5.71 in the fourth layer.
Correlation-based re-alignment (A6) is also effective in reducing the misalignment, particularly in the lower layers (see Fig.~\ref{fig:visualise_misalignment}d 
for a visualisation of re-alignment results using this approach).

We observe an interesting phenomenon in the third layer.
The misalignment is only slightly reduced, from 30.98 to 29.35 using bottom-up weight-based re-alignment (Table~\ref{table:misalignment_scores_alignment}).
Yet this layer does become more aligned, since activation-based misalignment metrics improve significantly (see Appendix~\ref{appendix:alignment_other_metrics}). 
We attribute the small reduction in misalignment to the third layer being large (625.5K weights) and placed in the middle of the network, making it sensitive to re-alignment errors propagated from the other layers.

\begin{table*}[!ht]
\centering
\caption{Parameters used in the experiments. We also report the accuracy of the target models.}
\resizebox{\textwidth}{!}{
\begin{tabular}{ccccccccccccc}
\toprule
Dataset & Architecture & Train acc. & Test acc. & $|D|$ &  $N_{\text{V}}$ & $|D_A|$ & $|D_{\text{target}}|$ & $D_A \cap D_{\text{target}}$ & $|D_T|$  & $N_{\text{train}}$ & $N_{\text{val}}$ & $N_{\text{test}}$ \\
\midrule
CIFAR10 &  CNN4 & 99.9 (0.1) & 67.5 (0.6) & 60000 & 2000 & 28000 & 28000 & $\emptyset$ & 14000 & 21000 & 2000 & 5000 \\
CIFAR10 & VGG16 & 100.0 (0.0) & 82.6 (0.4) & 60000 & 2000 & 56000 & 56000 & $D_A=D_{\text{target}}$ & 50000 & 8000 & 2000 & 2000 \\
Texas100 & MLP5 & 98.8 (2.7) & 48.7 (0.7) & 44000 & 2000 & 20000 & 20000 & $\emptyset$ & 10000 & 13000 & 2000 & 5000 \\
Purchase100 & MLP4 & 100.0 (0.0) & 80.9 (0.7) & 84000 & 2000 & 40000 & 40000 & $\emptyset$ & 20000 & 33000 & 2000 & 5000 \\
TinyImagenet200 & ResNet18 & 63.0 (1.0) & 32.3 (0.6) & 100000 & 2000 & 96000 & 96000 & $D_A=D_{\text{target}}$ & 86000 & 14000 & 2000 & 4000 \\
\bottomrule
\end{tabular}
}
\label{table:details}
\end{table*}

The weight sorting approach   (A1) of Ganju et al.~\cite{ganju2018property} performs poorly, on par with and sometimes slightly better than the no re-alignment baseline.
We attribute its poor performance to the neuron sorting criterion: the sum of weights may be too coarse to allow for effective re-alignment. 
More generally, canonical representations have been argued to be sensitive to small changes in weights~\cite{ashmore2015method}.

Finally, the re-alignment after initialisation technique (A2) proves to be ineffective, likely due to the inherent noise of matching the yet to be trained shadow model with the fully trained target model.

\textbf{Takeaways.}
The consistent improvements of re-alignment techniques (A3)-(A6) confirm that it is possible to reduce the  misalignment between the target model and the shadow models.
However, re-alignment techniques are imperfect and struggle with re-aligning large and middle layers.

%% file: sections/membership_inference.tex
\section{White-box membership inference}
\label{sec:membership_inference}
A final important criterion to judge the effectiveness of re-alignment techniques is their performance on white-box inference attacks.
In this section, we focus on white-box MIAs as a case study. 
Note that by definition of symmetries (Definition~\ref{def:symmetry}), re-alignment techniques do not modify the outputs of shadow models and hence cannot impact black-box MIAs.

To understand the impact of misalignment, we evaluate attacks relying on a membership meta-classifier (MC) trained on features derived either (S1) from the target model or  (S3) from shadow models trained on a different dataset and using a different randomness.
Under (S1), referred to as the auditor, the MC is trained on features computed on the target model and known training/unseen records $\mathcal{F}(\mathcal{M}_T, x, y), (x,y)\in D_{\text{in}}\cup D_{\text{out}}$.
There is no misalignment with respect to the features used at test time $\mathcal{F}(\mathcal{M}_T, x_T, y_T)$, which are also computed on the target model.
Under (S3), the MC is trained on features computed on shadow models $\mathcal{F}(\mathcal{M}_k, x, y), (x,y)\in D_{A}$, with $\mathcal{M}_k=\mathcal{T}(D_k,seed_k)$ and $seed_k\neq seed_T$.
Thus, there is misalignment with respect to the features used at test time $\mathcal{F}(\mathcal{M}_T, x_T, y_T)$.

We consider an additional scenario (S2) where the adversary knows $seed_T^{WI}$, i.e., can train shadow models having the same weight initialisation as the target model $\mathcal{M}_T$.
This adversary is informed by the findings of Sec.~\ref{sec:identifying_causes} that weight initialisation is the main reason for shadow model misalignment and that shadow models with same WI are more aligned to $\mathcal{M}_T$.
This adversary is relevant in model fine-tuning (where the base model is publicly available) or update settings.

\textbf{Put informally to gain intuition, if the MC is trained on features extracted from the same number of records under (S1), (S2) and (S3), then (S1) upper bounds the performance of (S2) and (S3).}
However, in our experiments, we extract features from several shadow models, meaning that the MC is trained on more data under (S2) and (S3), which is why in some cases we will see, e.g., (S2) performing slightly better than (S1).

\subsection{Evaluation setup}\label{subsec:evaluation_setup}

In each setup, we train a meta-classifier (MC) to perform the membership inference task.
We consider three types of features: (1) output activations (OA) $x^l$ computed at a given layer $l$, (2) gradients (G) of the loss function computed on the record $(x, y)$ with respect to the weights of a layer $l$: $\partial \mathcal{L}/\partial W^l_{i,j}(x, y), i \in \{1, \ldots, D^l \}, j \in \{1, \ldots, D^{l-1} \}$ (and corresponding biases), and (3) input activations $(W^L_{y, i}  x^{L-1}_i)_{i=1, \ldots, D^{L-1}}$ entering the neuron that encodes the ground-truth class $y$. 
(3) are  a finer grained version of the signal encoded by the model confidence for class $y$. 
(1) and (2) were first proposed by Nasr et al.~\cite{nasr2019comprehensive}, while (3) have been used by Leino et al.~\cite{leino2020stolen} to learn a displacement function with respect to the same features computed on a proxy model.
Our use of (3) is different and simpler, as we combine these features directly with (1) and (2) and show them to improve MIAs.
We describe in Appendix~\ref{appendix:meta-classiffier} the meta-classifier (MC) architecture and how we train it.

Given a dataset $D$, e.g., CIFAR10, we first set aside two disjoint validation sets $V_1$ and $V_2$  of size $N_V$ that will be used as validation data when training target and shadow models, respectively.
Of the remaining records, we sample two datasets $D_A$ (adversary's auxiliary knowledge) and $D_{\text{test}}$ on which the shadow models, resp. target models will be trained and tested.
We sample the target dataset $D_T$ randomly from $D_{\text{test}}$, and $D_1, \ldots, D_K$ randomly from $D_{A}$ to have the same size as $D_T$.
The most stringent attack scenario is $D_A \cap D_{\text{test}}=\emptyset$, which is useful in modelling a weak adversary.
However, this scenario limits the size of the training dataset, leading to more overfitted models.
To experiment with larger datasets, we will also consider $D_A = D_{\text{target}}=D \setminus (V_1 \cup V_2)$.
The meta-classifier training, validation and test datasets are of size $N_{\text{train}}$, $N_{\text{val}}$, and $N_{\text{target}}$ and are balanced with respect to the membership task.
We report in Table~\ref{table:details} the values used for each dataset. 

\subsection{Datasets and models}\label{subsec:datasets_models}
We evaluate the attacks on three real-world datasets commonly used to evaluate privacy risks~\cite{shokri2017membership,salem2018ml,nasr2019comprehensive}.

\textbf{CIFAR10}~\cite{krizhevsky2009learning}. This dataset is described in Sec.~\ref{subsec:misalignment_experimental_setup}.
We train two architectures on this dataset: the CNN network described in Sec.~\ref{subsec:misalignment_experimental_setup} (referred to as CNN4), and a VGG16 network.
For the latter, we follow the original implementation~\cite{simonyan2014very}, except that we use smaller hidden layer sizes for the classification head (512 instead of 4096) for computational reasons.
The training hyperparameters are the same as for CNN4 (Sec.~\ref{subsec:misalignment_experimental_setup}), except that we use a batch size of 128 and early stopping with a minimum learning rate of $10^{-3}$.

\textbf{Texas100}~\cite{datasets}
consists of 67330 records and 6169 attributes, labeled with one of 100 classes.
We randomly select 44000 records as the dataset $D$.
We train the 5-layer MLP from Nasr et al.~\cite{nasr2019comprehensive} (referred to as MLP5), having layer sizes of 1024, 512, 256, 128, 100. 
The hidden layers are followed by ReLU and dropout layers with $p=5\%$.
The other training hyperparameters are the same as for CNN4 (Sec.~\ref{subsec:misalignment_experimental_setup}), except that we use a learning rate of 0.001.

\textbf{Purchase100}~\cite{datasets} consists of 197324 records and 600 attributes, labeled with one of 100 classes.
We randomly select 84000 records as the dataset $D$.
We train a 4-layer MLP (referred to as MLP4) having layer sizes of 512, 256, 128, 100.
The hidden layers are followed by ReLU.
The other training hyperparameters are the same as for CNN4 (Sec.~\ref{subsec:misalignment_experimental_setup}), except that we use a learning rate of 0.001.

\textbf{Tiny-ImageNet-200}~\cite{le2015tiny} consists of 100000 $64\times64$ images, labeled with one of 200 classes.
We train a ResNet18 architecture~\cite{he2016deep} on this dataset, using a classification layer with an output size of 200 classes.
For computational reasons and to limit overfitting, we train this model for 6 epochs when it roughly converges (training it further would increase train accuracy without improving validation accuracy).
The other training hyperparameters are the same as for CNN4 (Sec.~\ref{subsec:misalignment_experimental_setup}), except that we use a learning rate of 0.005.

\begin{table*}[!ht]
\centering
\caption{CIFAR10 (CNN4): Results of white-box MIA using output activation (OA) or gradient (G) features from different layers (mean AUC with 95\% confidence interval). 
}
\resizebox{\textwidth}{!}{
\begin{tabular}{l|c|c|c|c|c}
\toprule
\multirow{2}{*}{\textbf{Source of meta-classifier training features}} & \textbf{Last layer} & \multicolumn{2}{c|}{\textbf{Second to last layer}} & \textbf{Third to last layer} & \textbf{Fourth to last layer}\\
 & G & OA & G & G & G \\
\midrule
(S1) Target model (auditor) & 0.837 $\pm$ 0.012 & 0.805 $\pm$ 0.014 & 0.810 $\pm$ 0.017 & 0.820 $\pm$ 0.016 & 0.834 $\pm$ 0.010 \\
\midrule
(S2) Shadow models (same WI) & 0.822 $\pm$ 0.012 & 0.750 $\pm$ 0.012 & 0.823 $\pm$ 0.015 & 0.811 $\pm$ 0.015 & 0.830 $\pm$ 0.009 \\
\midrule
(S3) Shadow models (all seeds $\neq$)  & 0.789 $\pm$ 0.021 & 0.509 $\pm$ 0.005 & 0.806 $\pm$ 0.014 & 0.788 $\pm$ 0.019 & 0.768 $\pm$ 0.024 \\
(S4) Shadow models (all seeds $\neq$) + weight sorting~\cite{ganju2018property} & 0.789 $\pm$ 0.025 & 0.511 $\pm$ 0.006 & 0.803 $\pm$ 0.013 & 0.791 $\pm$ 0.020 & 0.770 $\pm$ 0.023 \\
(S5) Shadow models (all seeds $\neq$) + bottom-up weight re-al. & 0.795 $\pm$ 0.017 & 0.581 $\pm$ 0.029 & 0.818 $\pm$ 0.017 & 0.802 $\pm$ 0.019 & 0.789 $\pm$ 0.022 \\
(S6) Shadow models (all seeds $\neq$) + top-down weight re-al. & \textbf{0.834} $\pm$ 0.013 & \textbf{0.813} $\pm$ 0.010 & \textbf{0.822} $\pm$ 0.019 & 0.800 $\pm$ 0.020 & 0.787 $\pm$ 0.021 \\
(S7) Shadow models (all seeds $\neq$) + activation re-al. & 0.797 $\pm$ 0.023 & 0.652 $\pm$ 0.007 & 0.820 $\pm$ 0.015 & \textbf{0.803} $\pm$ 0.018 & \textbf{0.793} $\pm$ 0.024 \\
(S8) Shadow models (all seeds $\neq$) + correlation re-al. & 0.795 $\pm$ 0.013 & 0.639 $\pm$ 0.014 & 0.818 $\pm$ 0.017 & 0.803 $\pm$ 0.019 & 0.796 $\pm$ 0.021 \\
(S9) Shadow models (all seeds $\neq$) + re-aligned after init. & 0.818 $\pm$ 0.014 & 0.757 $\pm$ 0.015 & 0.809 $\pm$ 0.018 & 0.786 $\pm$ 0.019 & 0.768 $\pm$ 0.023 \\
\bottomrule
\end{tabular}
}
\label{table:mia_layer_by_layer_cnn4}
\end{table*}

\begin{table*}[!ht]
\centering
\caption{CIFAR10 (VGG16): Results of white-box MIAs trained on output activation (OA) or gradient (G) features from different layers (mean AUC with 95\% confidence interval). 
}
\resizebox{\textwidth}{!}{
\begin{tabular}{l|c|c|c|c|c}
\toprule
\multirow{2}{*}{\textbf{Source of meta-classifier training features}} & \textbf{Last layer} & \multicolumn{2}{c|}{\textbf{Second to last layer}} & \multicolumn{2}{c}{\textbf{Third to last layer}} \\
 & G & OA & G & OA & G \\
\midrule
(S1) Target model (auditor) & 0.655 $\pm$ 0.005 & 0.686 $\pm$ 0.005  & 0.665 $\pm$ 0.007 & 0.682 $\pm$ 0.006 & 0.672 $\pm$ 0.006 \\
\midrule
(S2) Shadow models (same WI) & 0.644 $\pm$ 0.004 & 0.670 $\pm$ 0.008 & 0.648 $\pm$ 0.006 & 0.651 $\pm$ 0.011 & 0.648 $\pm$ 0.006 \\
\midrule
(S3) Shadow models (all seeds $\neq$) & 0.643 $\pm$ 0.006 & 0.525 $\pm$ 0.006 & 0.642 $\pm$ 0.006 & 0.501 $\pm$ 0.012 & 0.648 $\pm$ 0.004 \\
(S5) Shadow models (all seeds $\neq$) + bottom-up weight re-al. & 0.644 $\pm$ 0.008 & 0.530 $\pm$ 0.013 & 0.645 $\pm$ 0.004 & 0.528 $\pm$ 0.012 & 0.654 $\pm$ 0.004 \\
(S6) Shadow models (all seeds $\neq$) + top-down weight re-al. & 0.647 $\pm$ 0.005 & 0.670 $\pm$ 0.008 & 0.647 $\pm$ 0.005 & 0.657 $\pm$ 0.010 & 0.657 $\pm$ 0.006 \\
(S7) Shadow models (all seeds $\neq$) + activation re-al.  & 0.640 $\pm$ 0.006 & \textbf{0.672} $\pm$ 0.008 & 0.648 $\pm$ 0.004 & \textbf{0.673} $\pm$ 0.010 & 0.654 $\pm$ 0.004 \\
\bottomrule
\end{tabular}
}
\label{table:mia_layer_by_layer_vgg}
\end{table*}

\subsection{Results}\label{subsec:results_mia}
We train $R=10$ different target models with different weight initialisation (WI) on the same dataset $D_T$.
For each target model, we train $K=10$ shadow models with the same WI as the target model for scenario (S2) and $K=10$ shadow models with a different WI for (S3).
As an optimisation, under (S3) we reuse the same shadow models for all target models, ensuring their seeds are different from all the target models' seeds.
For Tiny-ImageNet-200 and ResNet18, due to computational constraints, we only run $R=5$ repetitions of the experiment using $K=2$ shadow models for each target model.
As before, under (S3) we reuse the same $K=2$ shadow models across target models.
In each scenario, we train the MC and report its Area Under the Curve (AUC) computed on the $N_{\text{test}}$ records (mean and 95\% confidence interval over the $R$ runs).

Table~\ref{table:mia_layer_by_layer_cnn4} illustrates the impact of misalignment on white-box MIAs against CNN4.
When the features used are the output activations (OA) of the second to last layer, the AUC drops from 0.805 for the auditor (S1) to 0.750 for the same weight initialisation (WI) adversary (S2).
Next, lacking knowledge of the WI, the classical adversary (S3) training the meta-classifier (MC) on features derived from misaligned shadow models achieves random performance, at 0.509.
However, the top-down weight matching technique (S6) is able to fully recover the signal, reaching 0.813.
Except weight sorting~\cite{ganju2018property}, all other methods improve the AUC of MIAs.

Table~\ref{table:mia_layer_by_layer_vgg} shows similar results on VGG16.
We observe a slight drop in AUC from 0.686 (S1) to 0.670 (S2), followed by a large drop to 0.525 (S3) for MIAs using the OAs of the second to last layer.
Similarly, the AUC slightly drops from 0.682 (S1) to 0.651 (S2) and then to random performance: 0.501 (S3) for MIAs using the OAs of the third to last layer.
The re-alignment techniques are able to restore the performance of OAs, to 0.672 (S7) in the second to last layer and 0.673 (S7) in the third to last layer.

\textbf{Takeaway.} When a model is only partially released, without the last layer, e.g., for embedding purposes, the adversary cannot compute gradients, only activations.
Our results strongly suggest that when the OAs of lower layers contain MIA signal, this signal can be recovered by the re-aligned shadow models, while MIAs using misaligned shadow models would perform very poorly.

\begin{table*}[!ht]
\centering
\caption{MIAs using different combinations of features extracted from the last layer, in the auditor (S1) scenario. The best results are highlighted in bold.}
\begin{tabular}{l|ccc|c}
\toprule
\multirow{2}{*}{\textbf{Dataset}} & \multicolumn{3}{c|}{\textbf{White-box}} &  \textbf{Black-box} \\
 & OA + IA + G (Ours) & OA + G (Nasr et al.~\cite{nasr2019comprehensive}) & OA + IA (Ablation) & OA (Shokri et al.~\cite{shokri2017membership}) \\
 \midrule
 CIFAR10 (CNN4) & \textbf{0.860} $\pm$ 0.013 & 0.839 $\pm$ 0.012 & 0.846 $\pm$ 0.010 & 0.816 $\pm$ 0.010 \\
 CIFAR10 (VGG16) & \textbf{0.685} $\pm$ 0.005  & 0.675 $\pm$ 0.004 & 0.680 $\pm$ 0.006 & 0.637 $\pm$ 0.014 \\
 Texas100 (MLP5) & \textbf{0.912} $\pm$ 0.020 & 0.904 $\pm$ 0.021 & 0.907 $\pm$ 0.019 & 0.880 $\pm$ 0.015 \\
 Purchase100 (MLP4) & \textbf{0.757} $\pm$ 0.013 & 0.739 $\pm$ 0.012 & 0.739 $\pm$ 0.014 & 0.733 $\pm$ 0.010 \\
 Tiny-ImageNet-200 (Resnet18) & \textbf{0.693} $\pm$ 0.008 & 0.685 $\pm$ 0.005  & 0.687 $\pm$ 0.010 & 0.597 $\pm$ 0.010 \\
\bottomrule
\end{tabular}
\label{table:results_ia}
\end{table*}

\begin{table*}[!ht]
\centering
\caption{CIFAR10 (CNN4): Results of white-box MIAs using features from multiple layers.
}
\begin{tabular}{l|c|ccc}
\toprule
\multirow{2}{*}{\textbf{Source of meta-classifier training features}} & \textbf{Black-box} & \multicolumn{3}{c}{\textbf{White-box}} \\
\cmidrule{2-5}
& Last layer  & Last layer & Last two layers & All four layers \\
\midrule
(S1) Target model (auditor)  & 0.816 $\pm$ 0.010 & 0.860 $\pm$ 0.013 & 0.857 $\pm$ 0.014 & 0.853 $\pm$ 0.015 \\
\midrule
(S2) Shadow models (same WI) & 0.810 $\pm$ 0.011 & 0.857 $\pm$ 0.013 & 0.859 $\pm$ 0.019 & 0.862 $\pm$ 0.018 \\
\midrule
(S3) Shadow models (all seeds $\neq$) & 0.815 $\pm$ 0.011 & 0.837 $\pm$ 0.014 & 0.837 $\pm$ 0.015 & 0.842 $\pm$ 0.015 \\
(S4) Shadow models (all seeds $\neq$) + weight sorting~\cite{ganju2018property} & 0.813 $\pm$ 0.012 & 0.838 $\pm$ 0.014 & 0.840 $\pm$ 0.015 & 0.844 $\pm$ 0.015 \\
(S5) Shadow models (all seeds $\neq$) + bottom-up weight re-al. & 0.813 $\pm$ 0.012 & 0.853 $\pm$ 0.014 & 0.850 $\pm$ 0.018 & 0.855 $\pm$ 0.018 \\
(S6) Shadow models (all seeds $\neq$) + top-down weight re-al. & 0.815 $\pm$ 0.013 & 0.849 $\pm$ 0.017 & 0.851 $\pm$ 0.017 & 0.856 $\pm$ 0.019 \\
(S7) Shadow models (all seeds $\neq$) + activation re-al. & 0.815 $\pm$ 0.012 & \textbf{0.860} $\pm$ 0.014 & 0.861 $\pm$ 0.017 & 0.860 $\pm$ 0.017 \\
(S8) Shadow models (all seeds $\neq$) + correlation re-al. & 0.812 $\pm$ 0.010 & 0.858 $\pm$ 0.015 & \textbf{0.861} $\pm$ 0.015 & \textbf{0.863} $\pm$ 0.018 \\
(S9) Shadow models (all seeds $\neq$) + re-al. after init. & 0.811 $\pm$ 0.010 & 0.844 $\pm$ 0.015 & 0.843 $\pm$ 0.019 & 0.844 $\pm$ 0.016 \\
\bottomrule
\end{tabular}
\label{table:mia_cifar10}
\end{table*}

The gradients are also affected by misalignment, although much less, as we observe a small drop in AUC when moving from (S1) to (S3).
This is likely due to gradients of members and non-members being easier to distinguish due to differences in norm~\cite{nasr2019comprehensive}, something which the MC can learn in spite of misalignment.
On CNN4, we find that re-alignment techniques are able to recover the signal, e.g., top-down weight re-alignment reaches 0.834 (S6), much better than 0.789 using no re-alignment (S3), and comparable to the auditor (S1), at 0.837.
As expected, the MIA performance of the same weight initialisation adversary (S2) is less affected by  misalignment, reaching 0.822.
As for the first two layers, we observe similar trends, although none of the alignment techniques fully recovers the signal.
On VGG16, re-alignment techniques provide a small or no improvement to gradient-based MIAs.

We show in Table~\ref{table:results_ia} that input activations (IA) extracted from the last layer improve the performance of white-box MIAs consistently across datasets.
To minimise the effect of misalignment and isolate the improvement due to IAs alone, we perform a detailed ablation in the auditor setup (S1).
Our results suggest that IAs add new membership signal that is not available in the gradients.
On CIFAR10 for instance, the AUC increases from 0.816 (black-box, OA) to 0.860 (white-box, OA + IA + G), better than OA + G (white-box~\cite{nasr2019comprehensive}), at 0.839.
\textbf{In light of these results, we believe that input activations, which are simple and inexpensive to compute, should be incorporated in future ML privacy risks assessments.}

\begin{table*}[!ht]
\centering
\caption{Texas100 (MLP5): Results of white-box MIAs using features from the last layer.
}
\begin{tabular}{l|c|c}
\toprule
\textbf{Source of meta-classifier training features} & \textbf{Black-box (OA)} & \textbf{White-box (OA + IA + G)} \\
\midrule
(S1) Target model (auditor) & 0.880 $\pm$ 0.015 & 0.912 $\pm$ 0.020 \\
\midrule
(S2) Shadow models (same WI) & 0.843 $\pm$ 0.023 & 0.873 $\pm$ 0.041 \\
\midrule
(S3) Shadow models (all seeds $\neq$) & 0.859 $\pm$ 0.015 & 0.857 $\pm$ 0.048 \\
(S4) Shadow models (all seeds $\neq$) + weight sorting~\cite{ganju2018property} & -  & 0.866 $\pm$ 0.041 \\
(S5) Shadow models (all seeds $\neq$) + bottom-up weight re-al. & - &\textbf{0.878} $\pm$ 0.029 \\
(S6) Shadow models (all seeds $\neq$) + top-down weight re-al. & - & 0.876 $\pm$ 0.039 \\
(S7) Shadow models (all seeds $\neq$) + activation re-al. & - & 0.871 $\pm$ 0.045 \\
(S8) Shadow models (all seeds $\neq$) + correlation re-al. & - & 0.859 $\pm$ 0.050 \\
\bottomrule
\end{tabular}
\label{table:mia_texas100}
\end{table*}

Next, we show results of white-box MIAs that combine features from more layers, progressively adding new layers from top to bottom. 
This includes as a particular case the black-box attack of Shokri et al.~\cite{shokri2017membership} that uses the output activations (OA).
We find Shokri et al.'s attack to perform on par with the black-box LiRA attack of Carlini et al.~\cite{carlini2022membership} in our setting, where we only use 10 shadow models (see Appendix~\ref{appendix:lira} for details).
The white-box attacks combine the OAs and gradients (G) of each layer with the input activations (IA) of the last layer.

Table~\ref{table:mia_cifar10} shows that the performance of white-box MIAs on CNN4 is affected by shadow model misalignment, as it drops from 0.860 in the strong auditor scenario (S1) to 0.837 in classical adversary scenario (S3).
In contrast, the black-box MIA is not affected by shadow model misalignment, as its accuracy is roughly constant at 0.815.
The re-alignment techniques, and in particular bottom-up correlation-based re-alignment, are able to recover the lost signal, increasing the performance to e.g., 0.858 using features in the last layers and 0.863 using features in all the layers.
The performance of the same model initialisation adversary (S2) is similar to the auditor (S1), suggesting that \textbf{releasing the weight initialisation increases the risk of attacks}.
On the VGG16 network, we observe no drop between (S1) and (S2), and a small drop at (S3) for white-box attacks, with re-alignment techniques slightly improving the average performance, e.g., from 0.679 to 0.686.

Table~\ref{table:mia_texas100} shows that when applied to an MLP trained on Texas100, the classical and black-box adversaries perform similarly, reaching 0.857 and 0.859, respectively.
Re-aligning the shadow models however increases the performance of the classical adversary to 0.878 (bottom-up weight re-alignment) (S5), with similar results for top-down weight (S6) and activation (S7) re-alignment (0.876 and 0.871, respectively).
We observe a noticeable performance gap compared to the auditor (S1) (0.912).
We also noticed that the training procedure stopped much earlier for some shadow and target models compared to others.
We thus suspect the gap might be due to higher variability between models which might increase misalignment.

Tables~\ref{table:mia_cifar10_vgg16}- ~\ref{table:mia_resnet18} in Appendix~\ref{appendix:additional_results} reports attack results for VGG16, MLP, and ResNet18 architectures  trained on CIFAR10, Purchase100 and Tiny-ImageNet-200, respectively.
On VGG16 and Purchase100, the gap between the auditor (S1) and the classical adversary (S3) is smaller than in previous datasets (0.689 vs 0.679 and 0.757 vs 0.745, resp.), and the re-alignment techniques do not significantly improve the attack performance.
On Tiny-ImageNet-200, the gap is even smaller (0.690 vs 0.687) making the use of re-alignment less interesting (0.691).
We attribute these findings to the fact that while gradient features hold most of the MIA signal, they are, in these instances, not affected by misalignment.

\textbf{Worst-case privacy.} 
Carlini et al.~\cite{carlini2022membership} have argued that attacks should be considered successful whenever they are able to confidently target at least a few records, even if their accuracy is low.
We compute the complete Receiver Operator Characteristic (ROC) curves of attacks on a log-log scale, and compare the various re-alignment techniques according to the True Positive Rate (TPR) for low values of the False Positive Rate (FPR). 
The smallest FPR we consider is 1\% as using a smaller value, e.g., 0.1\% would lead to a noisy value of TPR.
We average the ROC curves over 10 experiments.
Fig.~\ref{fig:roc_curves}a in Appendix~\ref{appendix:additional_results} shows that on CIFAR10, our white-box approach using re-alignment achieves a TPR of 12.3\% for an FPR of 1\% and performs better than (1) the black-box attack of Shokri et al.~\cite{shokri2017membership} (7.2\%), (2) the white-box attacks with no re-alignment (7.8\%), and (3) the weight sorting technique of Ganju et al.~\cite{ganju2018property} (7.8\%).
Fig.~\ref{fig:roc_curves}b and c report similar results on Texas100 and Purchase100, although the differences are less marked.

\textbf{Comparison with set-based classifiers.}
An alternative approach to deal with the permutation equivalence symmetry, which was explored in prior works on property inference attacks (PIA)~\cite{ganju2018property,suri2022formalizing}, is the use of set-based classifiers.
A set-based classifier views the neurons in a layer as a set and aggregates their feature representations using a permutation-invariant operation such as summation. 
In PIAs, a set-based classifier is trained on the \textit{weights} $\theta$ of a model $f_{\theta}$ to infer a property of the training dataset, e.g., the proportion of females in the dataset.
We here extend set-based classifiers to white-box MIAs, using activations and gradients as features instead of the weights.
Our implementation is described in Appendix~\ref{appendix:set-based}.

Table~\ref{table:set_based_classifier}  shows that on the CIFAR10 dataset, the set-based classifier performs on par with the classical adversary (S3) without re-alignment and less well than our re-alignment techniques.
On Texas100, the set-based classifier performs similarly to our re-alignment techniques, recovering part of the MIA signal, but less well than the auditor setting (S1).
We attribute this to the difficulty of training set-based classifiers on large sets (here, neuron sets), as the signal coming from individual neurons is more ``smoothed'', i.e., averaged out.
Indeed, the CIFAR10 dataset has 500 neurons in the second to last layer, while Texas100 has only 128 neurons, making it harder to effectively train a set-based classifier in the former case.

\textbf{Comparison with Leino and Fredrikson~\cite{leino2020stolen}.} We further compare our approach with the Bayes-WB approach of ~\cite{leino2020stolen} (see Appendix~\ref{appendix:leino} for details).
Table~\ref{table:comparison_stolen_memories} in the Appendix shows that Bayes-WB performs worse than the (S3) adversary using the input activations (IA), reaching 0.708 and 0.726, resp. on CIFAR10.
Our top-down weight re-alignment technique (S6) achieves much better performance (0.805).
We conclude that although proxy models are designed to be ``aligned'' in some sense with the target model at the last layer, they do not replicate well the dataset memorisation patterns of the target model, likely due to only the last layer being trained on the dataset while the lower layers are maintained frozen.

\textbf{Impact of defenses.} We finally experiment with models trained with defenses: regularisation and differential privacy~\cite{dwork2006calibrating}.
It is indeed possible that defenses that are effective in the black-box setting remain vulnerable to white-box attacks leveraging new information.
Our results in Appendix~\ref{appendix:defenses} show that implementing these defenses reduces the risk of MIAs to close to random performance, albeit at a cost in accuracy.

%% file: sections/discussion.tex
\section{Future work and conclusion}
\label{sec:discussion}
In this paper, we identify shadow model misalignment as a  potential reason for the sub-optimal performance of white-box MIAs.
We systematically investigate its causes, extend and evaluate the effectiveness of re-alignment methods proposed in the model fusion literature to the shadow modelling context, and perform a comprehensive evaluation of white-box MIAs. 
We show that re-alignment techniques can improve MIA accuracy, sometimes by a large margin, while they come at no cost to the adversary. Indeed, they are inexpensive to run and across experiments, we do not find a single instance where re-alignment decreases performance compared to no re-alignment. This strongly suggests that, at worst, re-alignment does not affect the performance of attacks and at best improves performance. We conclude that they should be incorporated in privacy assessments of ML models, and more generally that MIAs should be designed taking into account the symmetries of DNNs. We now highlight potential avenues for extending our work.

Our work focuses on permutation equivalence-type symmetries, using standard re-alignment techniques from the model fusion literature. 
Future work could develop re-alignment techniques dealing with all known symmetries. 
As we observe from the distance metrics of the re-aligned models, there is scope for designing better re-alignment techniques that will further reduce the distance between the shadow models and the target model.
Future work could further analyse the symmetries present in other architectures, such as transformers.

The white-box MIAs considered in this work rely on a common meta-classifier for all the target records.
Targeted attacks using a record-dependent decision boundary have been shown to perform even better in the black-box setting~\cite{ye2022enhanced,carlini2022membership}.
We believe that targeted white-box attacks combined with our re-aligned shadow models are likely to further increase the risk.
A potential and yet untapped advantage is the use of weights as attack features.
Weights would be available since hundreds of models would typically be trained with or without the target record~\cite{carlini2022membership}.
Whether weights contain new membership signal that is not redundant with the activations and gradients is an open question.

%% file: sections/acknowledgments.tex
\begin{acks}
The authors thank Florimond Houssiau for his detailed feedback on the paper, Shubham Jain and Vince Guan for proofreading an earlier version, Lukas Wutschitz, Santiago Zanella-B\'eguellin, Victor R\"uhle, and Boris K\"opf for the useful discussions and feedback, and Robert Sim for help with releasing the code.
This work was supported by the PETRAS National Centre of Excellence for IoT Systems Cybersecurity, funded by the UK EPSRC under grant number EP/S035362/1.
\end{acks}

%% file: sections/appendix.tex
\appendix

\section{Appendix}
\label{sec:appendix}

\subsection{Misalignment trends using different hyperparameters and metrics}
\label{appendix:misalignment_metric}

\textbf{Other datasets.} Table~\ref{table:cifar100:misalignment_scores_causes} reports the weight misalignment scores (WMS)  using CNN4 but computed in the CIFAR100 dataset, with similar results as the ones obtained on the CIFAR10 dataset (Tables~\ref{table:misalignment_scores_causes} and ~\ref{table:misalignment_scores_alignment}).
Table~\ref{table:purchase100:misalignment_scores_causes} reports the WMS for MLP4 trained on the Purchase100 dataset.
We again observe a similar trend, but we notice that for the MLP, the gap between (a) models trained with a different batch ordering or dropout selection and (b) the target model is higher than for the CNN (Tables~\ref{table:misalignment_scores_causes} and ~\ref{table:cifar100:misalignment_scores_causes}).
We attribute this behaviour to  fully connected layers being  larger (in our experiments) than convolutional layers, making them  less stable to differences in randomness.

\begin{table*}[!htbp]
\centering
\caption{Weight misalignment scores for CNN4 trained on the CIFAR100 dataset.}
\begin{tabular}{lcccc}
\toprule
\multirow{2}{*}{\textbf{Difference w.r.t. target model $\mathcal{M}_T$}}  & \textbf{First layer} & \textbf{Second layer} & \textbf{Third layer} & \textbf{Fourth layer}\\
   & \textbf{(20 filters)} & \textbf{(50 filters)} & \textbf{(500 neurons)} & \textbf{(100 neurons)} \\
\midrule
\multicolumn{5}{c}{\textbf{Weight misalignment scores ($\downarrow$ indicates better alignment)}} \\
\midrule
Random permutation & 12.62 (0.53) & 17.28 (0.18) & 37.02 (0.03) & 24.75 (0.13) \\
\midrule
$\neq$ Weight initialisation (WI) & 12.82 (0.45) & 17.87 (0.48) & 38.11 (1.29) & 25.12 (0.60) \\
$\neq$ Batch ordering (BO) & 3.82 (0.30) & 9.95 (0.17) & 29.13 (0.58) & 16.83 (0.24) \\
$\neq$ Dropout selection (DS) &  3.52 (0.43) & 9.52 (0.49) & 28.33 (1.44) & 16.39 (0.59) \\
\midrule
Overlapping datasets & 4.24 (0.54) & 11.00 (0.48) & 31.21 (0.89) & 18.93 (0.43) \\
Disjoint datasets (DD) & 4.23 (0.03) & 11.43 (0.19) & 31.63 (0.37) & 19.43 (0.22) \\
\midrule
$\neq$ BO, $\neq$ DS, and DD & 4.36 (0.14) & 11.36 (0.39) & 31.78 (1.52) & 19.48 (0.68) \\
$\neq$ MI, $\neq$ BO, and $\neq$ DS & 12.54 (0.63) & 17.57 (0.58) & 37.56 (1.39) &  24.89 (0.77) \\
All $\neq$ ($\neq$ WI, $\neq$ BO, $\neq$ DS, and DD) & 12.90 (0.20) & 17.31 (0.29) & 36.78 (0.50) & 24.49 (0.23) \\
\midrule
(A3) Bottom-up weight-based re-alignment  & 8.46 (0.51) & 14.61 (0.22) & 34.70 (0.44) & 23.29 (0.33) \\
(A4) Top-down weight-based re-alignment & 9.37 (0.77) & 15.46 (0.39) & 36.07 (0.48) & 20.00 (0.24) \\
(A5) Activation-based re-alignment  & 8.80 (0.63) & 15.03 (0.24) & 35.84 (0.44) & 22.91 (0.22) \\
(A6) Correlation-based re-alignment~\cite{li2015convergent} & 8.88 (0.41) &  15.04 (0.25) & 35.85 (0.46) & 22.99 (0.24) \\
\bottomrule
\end{tabular}
\label{table:cifar100:misalignment_scores_causes}
\end{table*}

\begin{table*}[!htbp]
\centering
\caption{Weight misalignment scores for MLP4 trained on the Purchase100 dataset.}
\begin{tabular}{lcccc}
\toprule
\multirow{2}{*}{\textbf{Difference w.r.t. target model $\mathcal{M}_T$}}  & \textbf{First layer} & \textbf{Second layer} & \textbf{Third layer} & \textbf{Fourth layer}\\
   & \textbf{(512 neurons)} & \textbf{(256 neurons)} & \textbf{(128 neurons)} & \textbf{(100 neurons)} \\
\midrule
\multicolumn{5}{c}{\textbf{Weight misalignment scores ($\downarrow$ indicates better alignment)}} \\
\midrule
Random permutation & 57.18 (0.26) & 36.46 (0.14) & 17.79 (0.13) & 18.80 (0.19) \\
\midrule
$\neq$ Weight initialisation (WI) & 56.30 (0.82) & 35.95 (0.91) & 17.68 (0.42) & 19.72 (0.23) \\
$\neq$ Batch ordering (BO) & 47.61 (1.69) & 31.98 (1.01) & 13.82 (0.45) & 14.66 (0.64) \\
$\neq$ Dropout selection (DS) & 45.02 (0.72) & 30.86 (0.70) & 13.18 (0.28) & 13.83 (0.39) \\
\midrule
Overlapping datasets & 50.26 (0.67) & 32.25 90.56) & 13.89 (0.17) & 14.94 (0.29) \\
Disjoint datasets (DD) & 49.96 (1.28) & 32.31 (0.50) & 13.76 (0.36) & 14.93 (0.42) \\
\midrule
$\neq$ BO, $\neq$ DS, and DD & 49.17 (0.89) & 31.70 (0.43) & 13.67 (0.23) & 14.73 (0.32) \\
$\neq$ MI, $\neq$ BO, and $\neq$ DS & 57.24 (0.91) & 36.54 (0.65) & 17.74 (0.19) & 19.67 (0.30) \\
All $\neq$ ($\neq$ WI, $\neq$ BO, $\neq$ DS, and DD) & 56.67 (1.04) & 36.37 (0.78) & 17.81 (0.34) & 19.58 (0.47) \\
\midrule
(A3) Bottom-up weight-based re-alignment  & 47.26 (1.01) & 28.79 (0.78) & 14.36 (0.33) & 14.17 (0.41) \\
(A4) Top-down weight-based re-alignment & 49.91 (1.05) & 28.49 (0.75) & 14.33 (0.30) & 12.71 (0.30) \\
(A5) Activation-based re-alignment  & 48.84 (1.03) & 30.42 (0.70) & 15.11 (0.34) & 14.04 (0.37) \\
(A6) Correlation-based re-alignment~\cite{li2015convergent} & 48.80 (1.14) & 36.09 (1.12) & 17.58 (0.35) & 19.18 (0.71) \\
\bottomrule
\end{tabular}
\label{table:purchase100:misalignment_scores_causes}
\end{table*}

\textbf{Impact of hyperparameters.} 
To understand the impact of the training hyperparameters on the WMS, we retrain the CNN4 architecture on CIFAR10 using different hyperparameters for the target model (propagating them to the shadow models). 
More specifically, we study the impact of the learning rate $\eta$, as it is proportional to changes applied to the weights in each gradient update, and of the early stopping patience $E$, as larger values can lead to longer training and therefore increase model overfitting.
Table~\ref{table:wms_comparison_hyperparameters} shows that although numbers vary, the trends of WMS are similar to the ones identified previously.
To minimise the impact of overfitting, in this experiment we trained the models on 50000 records, the largest training dataset size considered in the work.
Note that changing the dataset sizes impacts the magnitude of the scores, but the trends remain the same.
For completeness, we report in Table~\ref{table:wms_comparison_dataset_sizes} the WMS for various training dataset sizes and illustrate in Fig.~\ref{fig:changing_dataset_size} the activation maps in the first convolutional layer. 

\begin{table*}[!htbp]
\centering
\caption{Weight misalignment scores on CNN4 trained on CIFAR10 subsets of different sizes. 
}
\begin{tabular}{clcccc}
\toprule
\multirow{2}{*}{\textbf{Training size}} & \textbf{Difference w.r.t.}  & \textbf{First layer} & \textbf{Second layer} & \textbf{Third layer} & \textbf{Fourth layer}\\
& \textbf{target model $\mathcal{M}_T$} & \textbf{(20 filters)} & \textbf{(50 filters)} & \textbf{(500 neurons)} & \textbf{(10 neurons)} \\
\midrule
& Random permutation & 11.81 (0.51) & 16.58 (0.15) & 31.87 (0.06) & 12.80 (0.66) \\
& $\neq$ Weight initialisation (WI) & 12.09 (0.34) & 16.24 (0.20) & 30.46 (0.45) & 12.54 (0.14) \\
12500 & $\neq$ Batch ordering (BO) & 3.42 (0.34) & 8.88 (0.26) & 21.51 (0.85) & 6.62 (0.15) \\
& $\neq$ Dropout selection (DS) & 3.52 (0.30) & 8.72 (0.23) & 21.25 (0.46) & 6.47 (0.16) \\
& $\neq$ WI, $\neq$ BO, and $\neq$ DS & 12.05 (0.43) & 16.30 (0.21) & 30.53 (0.34) & 12.65 (0.18) \\
\midrule 
& Random permutation & 13.94 (0.58) & 19.08 (0.29) & 35.49 (0.06) & 14.14 (0.74) \\
& $\neq$ Weight initialisation (WI) & 14.58 (0.32) & 19.56 (0.51) & 36.02 (1.30) & 14.25 (0.16) \\
25000 & $\neq$ Batch ordering (BO) & 5.62 (1.89) & 12.27 (0.53) & 28.66 (1.03) & 7.87 (0.15) \\
& $\neq$ Dropout selection (DS) & 5.63 (0.12) & 12.05 (0.23) & 28.64 (0.39) & 7.80 (0.11) \\
& $\neq$ WI, $\neq$ BO, and $\neq$ DS & 14.48 (0.32) & 19.39 (0.22) & 35.63 (0.31) & 14.31 (0.10) \\
\midrule 
 & Random permutation & 16.74 (0.52) & 24.50 (0.41) & 48.78 (0.07) & 15.60 (0.87) \\
& $\neq$ Weight initialisation (WI) & 17.23 (0.19) & 24.23 (0.19) & 47.19 (0.45) & 15.51 (0.13) \\
50000 & $\neq$ Batch ordering (BO) & 8.01 (0.77) & 16.66 (0.61) & 41.41 (1.60) & 9.29 (0.08) \\
& $\neq$ Dropout selection (DS) & 6.54 (1.25) & 15.41 (0.38) & 40.51 (0.71) & 9.19 (0.06) \\
& $\neq$ WI, $\neq$ BO, and $\neq$ DS & 17.33 (0.29) & 24.53 (0.37) & 48.09 (1.02) & 15.75 (0.20) \\
\bottomrule
\end{tabular}
\label{table:wms_comparison_dataset_sizes}
\end{table*}

\begin{table*}[!htbp]
\centering
\caption{Weight misalignment scores on CNN4 (CIFAR10) using various learning rates $\eta$ and early stopping patience $E$. 
}
\begin{tabular}{llcccc}
\toprule
& \textbf{Difference w.r.t.}  & \textbf{First layer} & \textbf{Second layer} & \textbf{Third layer} & \textbf{Fourth layer}\\
& \textbf{target model $\mathcal{M}_T$} & \textbf{(20 filters)} & \textbf{(50 filters)} & \textbf{(500 neurons)} & \textbf{(10 neurons)} \\
\midrule
 & Random permutation & 16.74 (0.52) & 24.50 (0.41) & 48.78 (0.07) & 15.60 (0.87) \\
$\eta=0.01, E=5$ &  $\neq$ Weight initialisation & 17.23 (0.19) & 24.23 (0.19) & 47.19 (0.45) & 15.51 (0.13) \\
& $\neq$ Batch ordering & 8.01 (0.77) & 16.66 (0.61) & 41.41 (1.60) & 9.29 (0.08) \\
\midrule 
 & Random permutation & 9.37 (0.34) & 11.19 (0.14) & 21.25 (0.03) & 9.31 (0.47) \\
$\eta=0.001, E=5$ & $\neq$ Weight initialisation & 9.86 (0.28) & 11.38 (0.15) & 21.31 (0.08) & 9.48 (0.13) \\
& $\neq$ Batch ordering & 1.02 (0.10) & 2.62 (0.08) & 5.97 (0.20) & 2.26 (0.05) \\
\midrule 
 & Random permutation & 10.27 (0.36) & 12.13 (0.14) & 22.04 (0.03) & 10.27 (0.51) \\
$\eta=0.001, E=10$ & $\neq$ Weight initialisation  & 10.96 (0.29) & 12.50 (0.14) & 22.25 (0.08) & 10.63 (0.07) \\
& $\neq$ Batch ordering & 1.24 (0.12) & 3.16 (0.09) & 7.75 (0.12) & 2.75 (0.07) \\
\bottomrule
\end{tabular}
\label{table:wms_comparison_hyperparameters}
\end{table*}

\textbf{Other metrics.} In addition to the WMS described in the main paper, we explore two other metrics computed using the output activation of a neuron (the 2-D activation map of a filter):
\begin{itemize}
\item \textbf{Activation misalignment score (AMS):}
In a layer  $l$ and a record $x$, the AMS is defined as $|| x^l(\mathcal{M}_T) -x^l(\mathcal{M})||_2$,
i.e., the Euclidean distance between the output activations computed on the target model and shadow model.
For an FC layer, $x^l$ is a vector of size $D^l$. 
For a convolutional layer, $x^l$ is a 3-D tensor, consisting of $C^l$ activation maps, that we flatten. 
We compute the average AMS over $R$ records.

\item \textbf{Correlation between activations (CBA).} We complement the two metrics with a third one~\cite{li2015convergent}, which measures how correlated neurons (filters) in the same position are across two models.
In an FC layer $l$, given a neuron position $d$, we consider the random variable $x_d^l$, where the randomness is taken over input records $x \sim D_{\text{A}}$.
We pass $R$ records through a model to obtain a series of $R$ activation values $S_d := ( (x_1)^l_d, \ldots, (x_R)^l_d )$.
We compute the CBA $\rho$ as the average (over all the positions $d$) of the Pearson correlation coefficient $\rho_d$ between the two series $S_d(\mathcal{M}_T)$ and $S_d(\mathcal{M})$: $\rho=(\sum_{d=1}^{D^l}\rho_d)/D^l$.
In a convolutional layer, activation maps replace the activation values.
We randomly select $P$ pixel coordinates from the activation map. 
We compute the Pearson correlation coefficient $\rho$ between the two series extracted from $\mathcal{M}$ and $\mathcal{M}_T$, respectively, for pixel $p$ and the activation map located in position $f$ (i.e., corresponding to the $f$-th filter).
We then compute the CBA as the average correlation over the $P$ pixels and $C^l$ filters.
\end{itemize}

The AMS ranges from 0 to $\infty$, is equal to 0 for two identical models, with lower values indicating a larger similarity between the internal representations of the models at layer $l$.
The CBA ranges between -1 and 1.
A value close to 0 indicates low correlation between the internal representations, while a value close to 1 indicates a high similarity.
Additionally, the correlation is scale-invariant, which allows to compare different layers.

We compute the activation misalignment score on $R=500$ validation records (the same for all the models).
We compute the correlation score on the same records and $P=50$ pixels.

Table~\ref{table:misalignment_scores_causes_other_metrics}A shows the AMS to follow the same trends as the WMS.
There is one exception: the fourth layer, whose output activations are always aligned, as they are trained to encode per-class probabilities.
For this reason, any difference between shadow models and $\mathcal{M}_T$ yields a much lower score compared to randomly permuting this layer.
The results on the fourth layer suggest that the input-output behaviours of shadow models deviate from the target model more when using $\neq$ WI (0.32) or a disjoint dataset (0.41) compared to applying other changes, e.g., $\neq$ BO (0.28).
We conclude that shadow models always learn a different classifier compared to the target model. 

Table~\ref{table:misalignment_scores_causes_other_metrics}B reports similar trends for the CBA.
We again observe that changing the weight initialisation has a very large impact on the correlation, across all internal layers.
This result explains the low AUC of the MIA reported in Sec.~\ref{sec:problem_statement}: the outputs of the third (second to last) layer in shadow models are uncorrelated with the target model ($\rho=0.01$, All $\neq$).
The positive correlation between the first layer and its permuted self ($\rho=0.15$), suggests the existence of redundancies in features learnt in the first layer.
The CBA metric also allows us to directly compare the levels of misalignment in different layers. 
For all the factors considered, the correlation decreases as we move from lower to upper layers.

\begin{table*}[!htbp]
\centering
\caption{Misalignment results using other metrics for CNN4 (CIFAR10). 
We highlight in bold the classical adversary (All $\neq$) and the adversary having knowledge of the target model's weight initialisation ($\neq$ BO, $\neq$ DS, and DD).
}
\begin{tabular}{lcccc}
\toprule
\multirow{2}{*}{\textbf{Difference w.r.t. target model} $\mathcal{M}_T$}  & \textbf{First layer} & \textbf{Second layer} & \textbf{Third layer} & \textbf{Fourth layer}\\
   & \textbf{(20 filters)} & \textbf{(50 filters)} & \textbf{(500 neurons)} & \textbf{(10 neurons)} \\
\midrule
\multicolumn{5}{c}{\textbf{A. Activation misalignment scores ($\downarrow$ indicates better alignment)}} \\
\midrule
Random permutation & 68.59 (3.10) & 72.70 (0.65) & 28.60 (0.05) & 1.02 (0.12) \\
\midrule
$\neq$ Weight initialisation (WI) & 68.31 (1.96) & 70.99 (1.46) & 27.34 (0.53) & 0.32 (0.01) \\
$\neq$ Batch ordering (BO) & 20.87 (2.21) & 37.37 (1.27) & 22.35 (0.50) & 0.28 (0.01) \\
$\neq$ Dropout selection (DS) & 21.88 (2.45) & 36.13 (1.01) & 21.88 (0.52) & 0.27 (0.01) \\
\midrule
Overlapping datasets & 25.28 (2.06) & 44.51 (0.67) & 23.73 (0.46) & 0.39 (0.01) \\
Disjoint datasets (DD) & 28.41 (2.79) & 46.49 (0.49) & 24.66 (0.35) & 0.41 (0.01) \\
\midrule
\textbf{$\neq$ BO, $\neq$ DS, and DD} & 30.54 (4.15) & 47.73 (1.40) & 24.85 (0.29) & 0.41 (0.02) \\
$\neq$ WI, $\neq$ BO, and $\neq$ DS & 68.90 (1.99) & 71.45 (1.72) & 27.59 (0.70) & 0.32 (0.01) \\
\textbf{All $\neq$ ($\neq$WI, $\neq$ BO, $\neq$ DS, and DD)}  & 68.64 (1.93) & 72.13 (0.29) & 27.90 (0.45) & 0.42 (0.03) \\
\midrule
\multicolumn{5}{c}{\textbf{B. Correlation between activations ($\uparrow$ indicates better alignment)}} \\
\midrule
Random permutation & 0.15 (0.07) & 0.04 (0.02) & 0.01 (0.00) & 0.04 (0.05)  \\
\midrule
$\neq$  Weight initialisation (WI) & 0.13 (0.03) & 0.03 (0.01) & 0.01 (0.00) & 0.28 (0.07) \\
$\neq$ Batch ordering (BO) & 0.74 (0.06) & 0.67 (0.02) & 0.28 (0.02) & 0.29 (0.07) \\
$\neq$ Dropout selection (DS) & 0.73 (0.04) & 0.68 (0.01) & 0.29 (0.01) & 0.29 (0.07) \\
\midrule
Overlapping datasets & 0.71 (0.06) & 0.56 (0.02) & 0.20 (0.02) & 0.26 (0.06) \\
Disjoint datasets (DD) & 0.71 (0.06) & 0.55 (0.02) & 0.18 (0.01) & 0.28 (0.07) \\
\midrule
\textbf{$\neq$ BO, $\neq$ DS, and $\neq$ DD} & 0.70 (0.05) & 0.53 (0.02) & 0.18 (0.01) & 0.28 (0.08) \\
$\neq$ WI, $\neq$ BO, and $\neq$ DS & 0.13 (0.03) & 0.04 (0.01) & 0.01 (0.00) & 0.28 (0.07) \\
\textbf{All $\neq$ ($\neq$WI, $\neq$ BO, $\neq$ DS, and DD)} & 0.14 (0.03) & 0.02 (0.01) & 0.01 (0.00) & 0.28 (0.08) \\
\bottomrule
\end{tabular}
\label{table:misalignment_scores_causes_other_metrics}
\end{table*}

\subsection{Visualisation of activation maps}
\label{appendix:visualisation}
\begin{figure*}[!htbp]
    \centering
    \subfigure[Target model $\mathcal{M}_T$]{
        \includegraphics[width=0.8\linewidth]{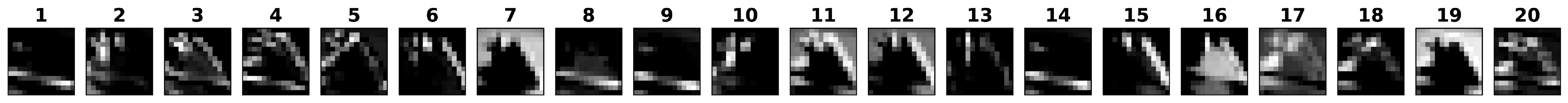}
    }
    \hfil
    \subfigure[Shadow models trained using different batch ordering]{
        \includegraphics[width=0.8\linewidth]{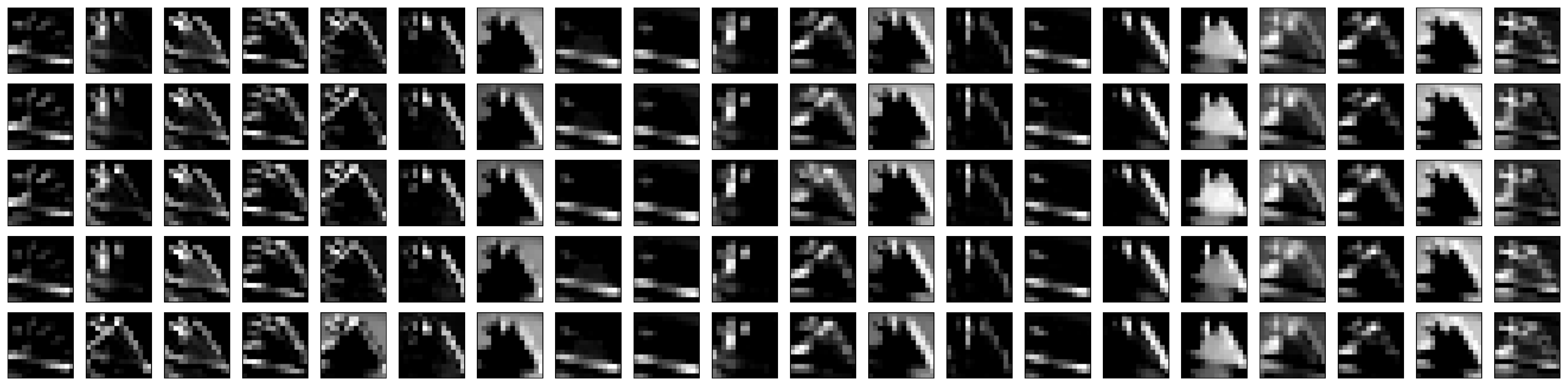} 
    }
    \hfil
    \subfigure[Shadow models trained using different dropout selection]{
        \includegraphics[width=0.8\linewidth]{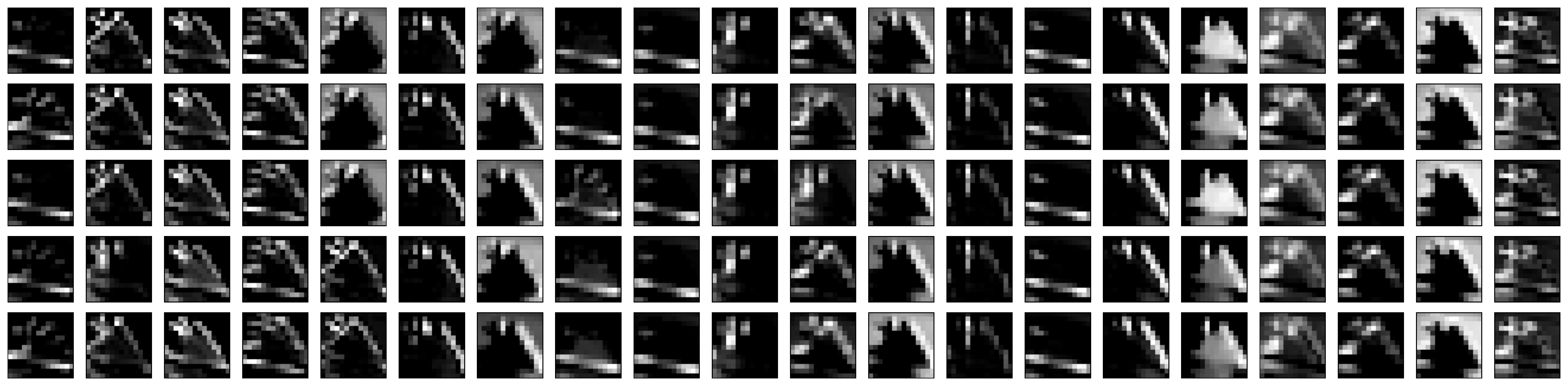} 
    }
    \hfil
    \subfigure[Shadow models trained using different weight initialisation]{
        \includegraphics[width=0.8\linewidth]{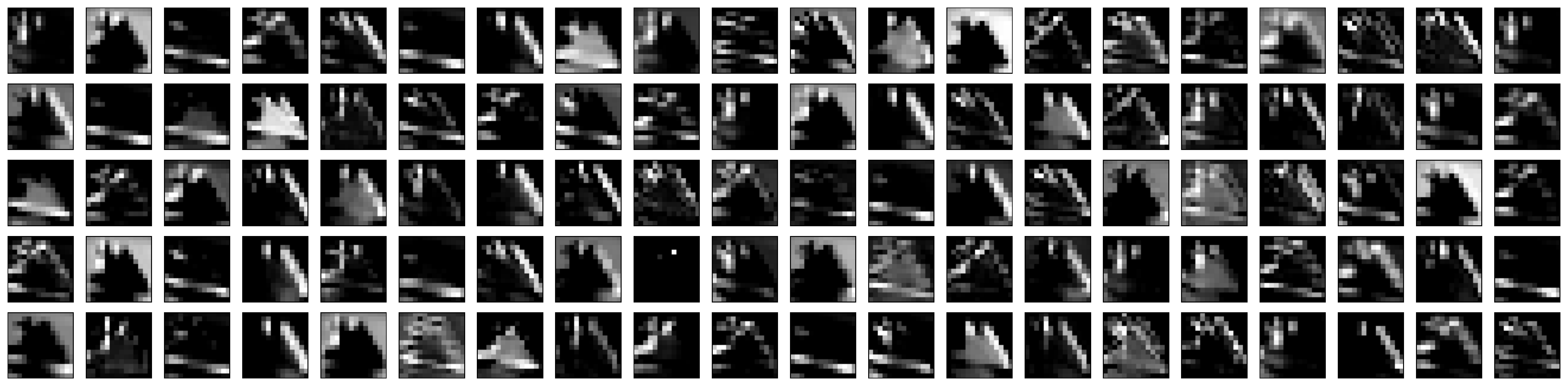} 
    }
    \caption{
    First layer activation maps for CNN4  models trained on CIFAR10, when changing the seeds controlling the training randomness.
    }\label{fig:changing_seed}
\end{figure*}

We consider, as before, the CNN4 architecture trained on the CIFAR10 dataset.
The figures below compare the activation maps of the first layer of the target model $\mathcal{M}_T$ with activation maps of:

\textbf{Fig. ~\ref{fig:changing_seed}:} CNN4 shadow models trained on the same dataset as the target model $\mathcal{M}_T$, but using a different seed for the batch ordering (Fig.~\ref{fig:changing_seed}b), dropout selection (Fig.~\ref{fig:changing_seed}c), and weight initialisation (Fig.~\ref{fig:changing_seed}d). 
Changing the weight initialisation results in mismatch between the features located in the same position across the models. 
Changing the seeds controlling the other sources of randomness does not, although some of the feature maps are slightly different.

\begin{figure*}[!ht]
    \centering
     \subfigure[Target model $\mathcal{M}^T$]{
        \includegraphics[width=0.8\linewidth]{images/target_model_12500_0-01.pdf}
    }
    \hfil
    \subfigure[Shadow models trained on overlapping datasets]{
        \includegraphics[width=0.8\linewidth]{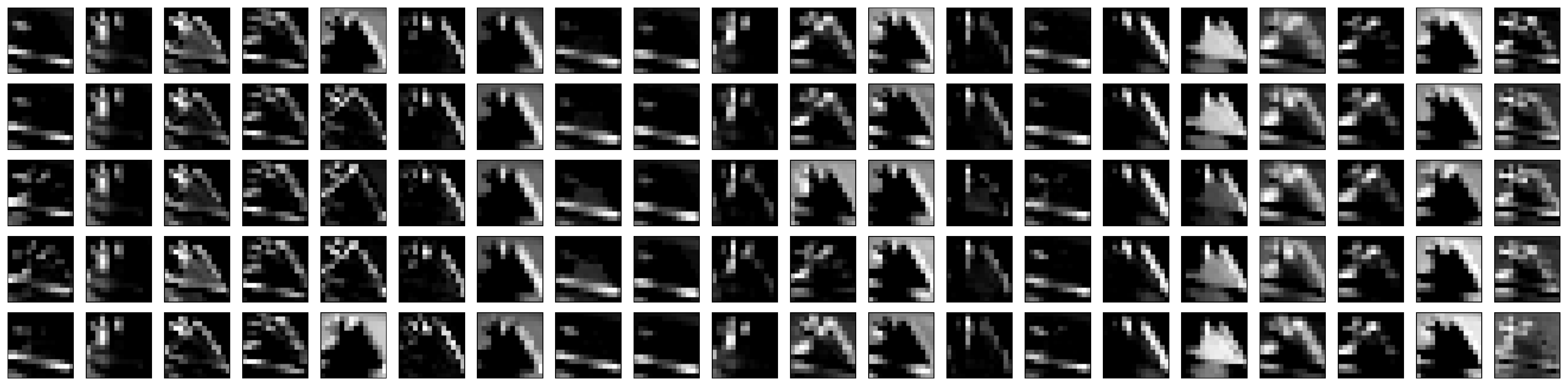} 
    }
    \hfil
    \subfigure[Shadow models trained on disjoint datasets]{
        \includegraphics[width=0.8\linewidth]{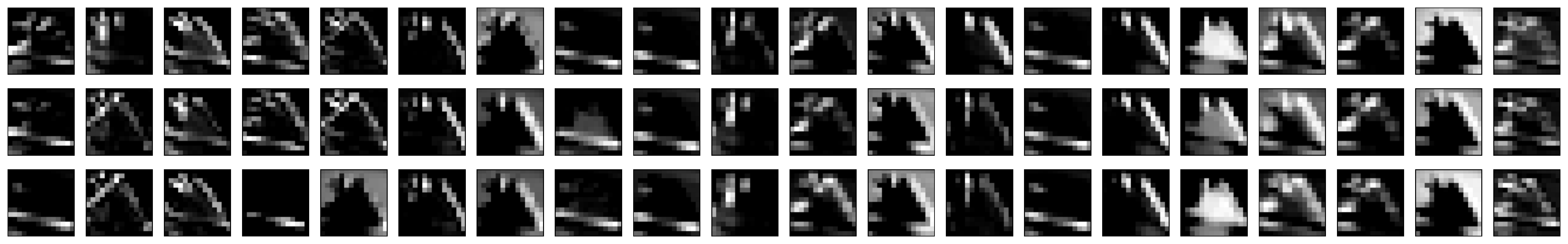} 
    }
    \caption{
    First layer activation maps for CNN4 models trained on different subsets of CIFAR10.}\label{fig:changing_dataset}
\end{figure*}

\textbf{Fig. ~\ref{fig:changing_dataset}:} CNN4 shadow models trained using the same weight initialisation as the target model $\mathcal{M}_T$, but on an overlappping (Fig.~\ref{fig:changing_dataset}b) or disjoint (Fig.~\ref{fig:changing_dataset}c) dataset.
Changing the dataset slightly impact the activation maps but does not result in misalignment, even when using disjoint datasets. 

\begin{figure*}[!htbp]
    \centering
    \subfigure[Target model $\mathcal{M}_T$]{
        \includegraphics[width=0.8\linewidth]{images/target_model_12500_0-01.pdf}
    }
    \hfil
    \subfigure[Shadow models trained using the same weight initialisation, but disjoint dataset and different batch ordering and dropout sampling seeds]{
        \includegraphics[width=0.8\linewidth]{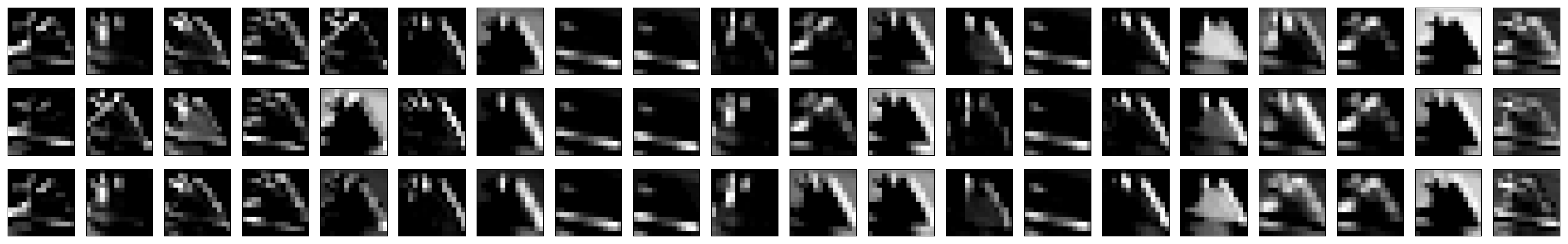}
    }
    \caption{First layer activation maps for CNN4 models trained on CIFAR10 using the same weight initialisation.
}\label{fig:changing_all_but_model_seed}
\end{figure*}

\textbf{Fig.~\ref{fig:changing_all_but_model_seed}:} CNN4 shadow models trained using different seeds as the target model $\mathcal{M}_T$ for the batch ordering, dropout selection, and a disjoint dataset, but the same weight initialisation (Fig.~\ref{fig:changing_all_but_model_seed}).
The activation maps are broadly (visually) aligned although we see some differences, e.g., in columns 1 and 7.

\begin{figure*}[!htbp]
     \centering
    \includegraphics[width=0.8\linewidth]{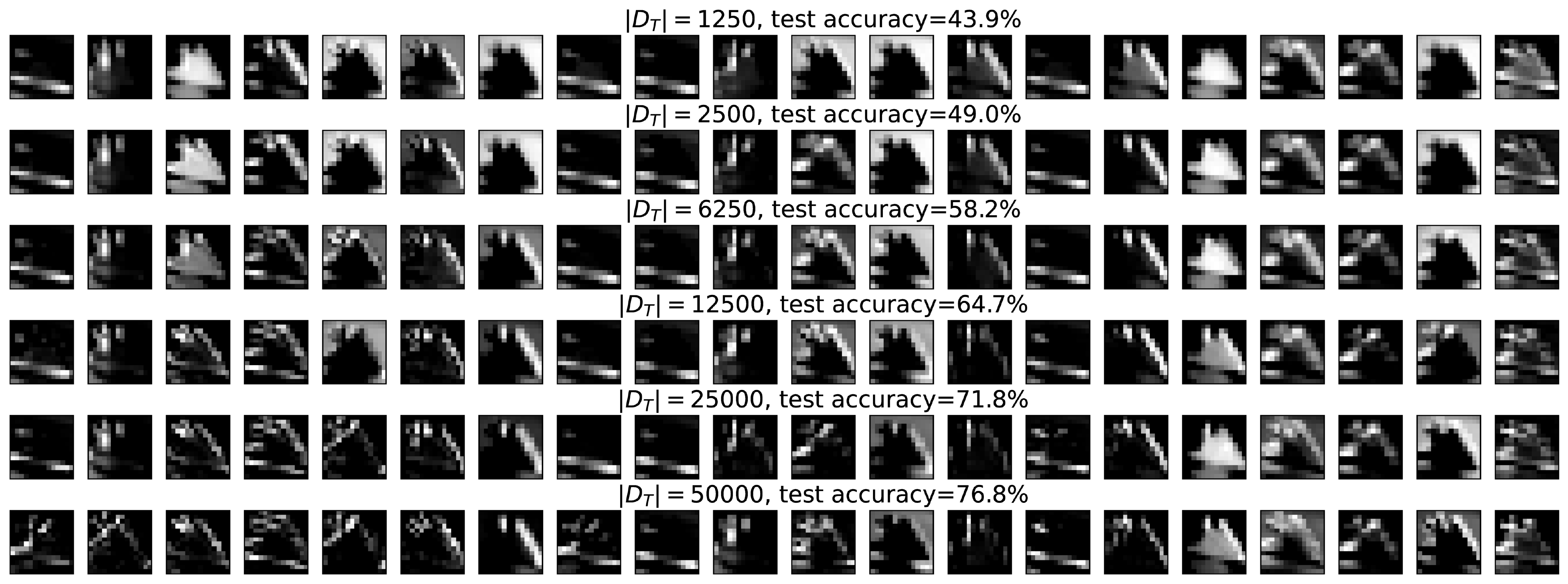} 
    \caption{
    First layer activation maps for CNN4 models trained on CIFAR10 subsets of different sizes.}\label{fig:changing_dataset_size}
\end{figure*}

\textbf{Fig~\ref{fig:changing_dataset_size}:} CNN4 shadow models trained on datasets of increasing sizes and same training randomness (Fig.~\ref{fig:changing_dataset_size}). 
Increasing the size of the dataset gradually changes the activation maps. As the model becomes less overfitted and learns more robust features, the activation maps stabilise and remain broadly (visually) aligned.

\subsection{Misalignment between models trained on different distributions}\label{appendix:celeba}
How does the adversary's knowledge about the target model's training distribution affect misalignment?
In a small experiment, we train shadow models to solve the same task as the target model, but on a different distribution. 
We use the CelebA dataset~\cite{liu2015faceattributes} containing 202599 faces of celebrities together with 40 attributes. 
We select smile prediction as the task and simulate two different distributions for the target and shadow models, respectively: 1) the target model's data distribution consists only of ``older'' faces (i.e., faces with negative label for the ``young'' attribute) and 2) the shadow models' data distribution consists of both ``young'' and ``older'' faces, in a proportion of 77.4/22.6\% as found in the CelebA dataset. This experiment simulates an adversary targeting a model trained on a user-specific distribution (e.g., an older person's device having pictures of similarly-aged individuals), such that the adversary does not know that the distribution only contains ``older'' faces. The adversary thus trains shadow models on a random distribution as found in the wild (e.g., in the CelebA dataset). 

We use the CNN4 architecture to train models on 20000 samples resized to $32\times 32$.
To simulate Adversary 1 who knows the target model's training distribution, we first train one shadow model on a subset of ``older'' faces disjoint from the target model's training dataset.
Note that we cannot train more than one shadow model on mutually disjoint subsets consisting of 100\% older faces, because CelebA contains only 45865 samples of older faces.
To simulate Adversary 2 who does not know the target model's training distribution, we train five shadow models on mutually disjoint subsets of faces from CelebA, such that the proportion of older faces roughly follows the naturally occurring proportion in the dataset (22.6\% older faces).
Both adversaries use different training randomness compared to the target model, where ``randomness'' encompasses all the seeds (weight initialisation, batch ordering, dropout sampling).

\begin{table*}[!htbp]
\centering
\caption{CelebA (CNN4): Weight misalignment scores for shadow models trained on the same/different distribution as the target model. The target model is trained on 100\% older faces.
}
\begin{tabular}{lcccc}
\toprule
\multirow{2}{*}{\textbf{Difference w.r.t. target model $\mathcal{M}_T$}}  & \textbf{First layer} & \textbf{Second layer} & \textbf{Third layer} & \textbf{Fourth layer}\\
   & \textbf{(20 filters)} & \textbf{(50 filters)} & \textbf{(500 neurons)} & \textbf{(10 neurons)} \\
\midrule
\multicolumn{5}{c}{\textbf{Weight misalignment scores ($\downarrow$ indicates better alignment)}} \\
\midrule
Random permutation & 6.93 (0.10) & 9.55 (0.09) & 19.09 (0.02) & 0.92 (2.06) \\
\midrule
(Adversary 1) Disjoint dataset of 100\% older faces, $\neq$ randomness & 6.89  & 10.09 & 19.19 & 3.34 \\
(Adversary 2) Random faces (22.6\% older), $\neq$ randomness & 6.88 (0.32) & 9.58 (0.18) & 19.07 (0.05) & 3.28 (0.11) \\
\midrule
(Ablation 1) Disjoint dataset of 100\% older faces, same randomness & 2.87 & 7.26 & 4.86 & 1.33 \\
(Ablation 2) Random faces (22.6\% older), same randomness & 2.96 (0.15) & 6.90 (0.19) & 4.51 (0.12) & 1.17 (0.10) \\
\bottomrule
\end{tabular}
\label{table:celeba}
\end{table*}

Table~\ref{table:celeba} shows similar WMS for Adversaries 1  and 2.
This suggests that shadow models trained on a different distribution w.r.t. the target model are not more misaligned to the target model than shadow models trained on the same distribution.

\begin{figure*}[!ht]
    \centering
    \subfigure[
    ]{
        \centering
        \includegraphics[scale=0.12]{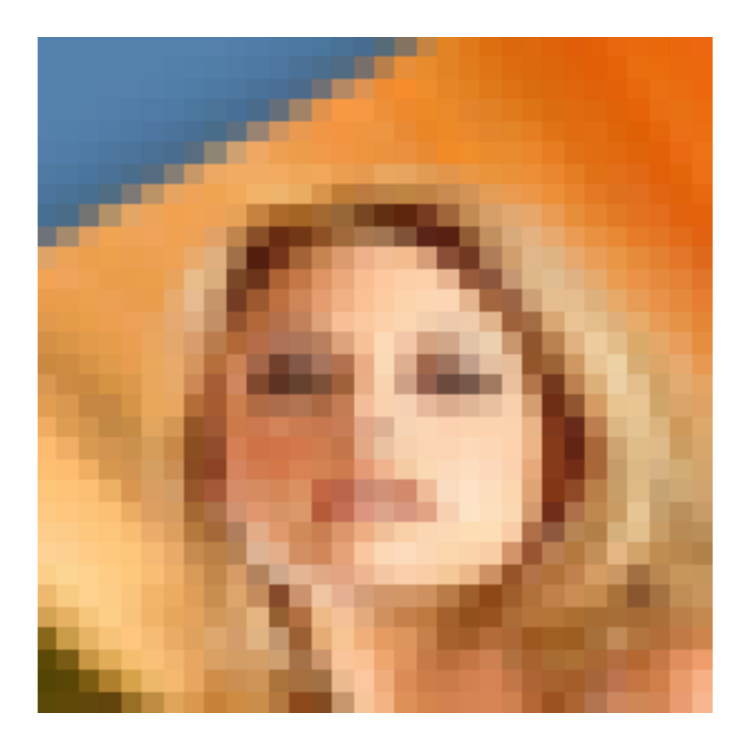} 
    }
    \hfil\\
    \subfigure{
        \includegraphics[width=0.8\linewidth]{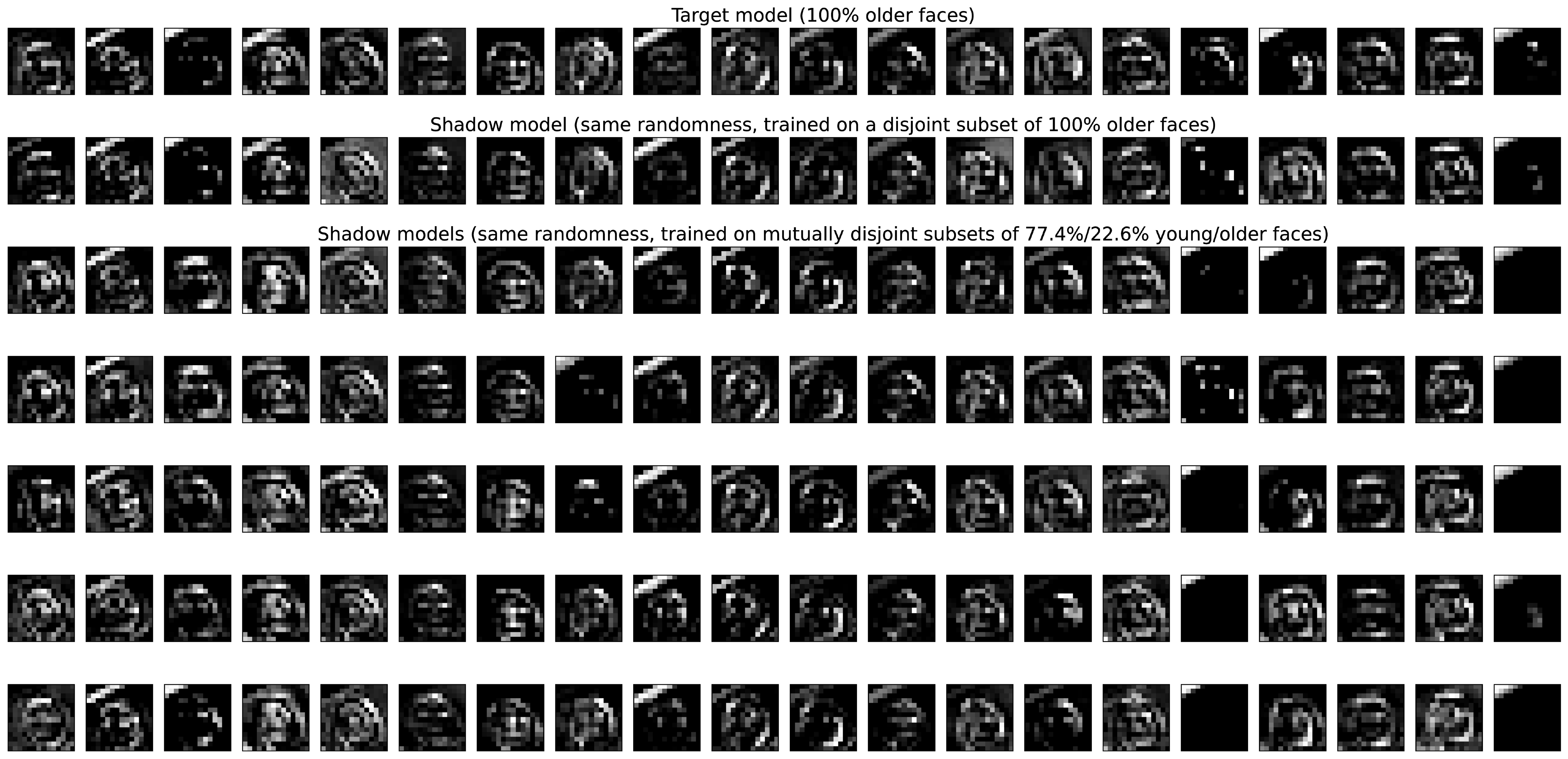}
    }
    \caption{First layer activation maps for CNN4 shadow models trained on CelebA subsets having the same/different distribution resp. to the target model. The activation maps are computed on a random image shown at the top (a).}\label{figure:celeba}
\end{figure*}

To understand whether the features learned from different distributions are different rather than just misaligned, we isolated the impact of the data distribution from the impact of the training randomness. More specifically, we use the same randomness for all the models and train shadow models either on datasets of the same distribution as the target model (Ablation 1) or from a different distribution (Ablation 2).
We do not observe a difference in misalignment between the two settings, i.e., we do not see the features of shadow models trained on predominantly young faces differ substantially from those of shadow models trained on 100\% older faces. Fig.~\ref{figure:celeba} confirms this with a visualisation.

\subsection{Re-alignment results for other metrics}
\label{appendix:alignment_other_metrics}
Table~\ref{table:misalignment_scores_alignment_other_metrics} reports the AMS and CBA for the CNN4 architecture trained on CIFAR10 after applying different re-alignment techniques.
We observe similar trends to the ones obtained using the weight-based misalignment score (Table~\ref{table:misalignment_scores_causes}).

\begin{table*}[!ht]
\centering
\caption{Misalignment results using other metrics when re-aligning CNN4 (CIFAR10) shadow models trained by the classical adversary.  
}
\begin{tabular}{lcccc}
\toprule
\multirow{2}{*}{\textbf{Re-alignment technique}}  & \textbf{First layer} & \textbf{Second layer} & \textbf{Third layer} & \textbf{Fourth layer}\\
 & \textbf{(20 filters)} & \textbf{(50 filters)} & \textbf{(500 neurons)} & \textbf{(10 neurons)} \\
\midrule
\multicolumn{5}{c}{\textbf{A. Activation misalignment scores ($\downarrow$ indicates better alignment)}} \\
\midrule
(A0) No re-alignment & 68.64 (1.93) & 72.13 (0.29) & 27.90 (0.45) & 0.42 (0.03) \\
\midrule
(A1) Weight sorting~\cite{ganju2018property} & 60.64 (0.73) & 70.19 (0.58) & 27.81 (0.36) & 0.42 (0.03) \\
(A2) Re-alignment after initialisation & 69.92 (1.64) & 70.66 (0.26) & 26.95 (0.25) & 0.42 (0.02) \\
(A3) Bottom-up weight-based re-alignment  & 43.22 (4.73) & 59.10 (0.82) & 25.79 (0.22) & 0.42 (0.03) \\
(A4) Top-down weight-based re-alignment & 48.07 (4.26) & 62.30 (0.81) & 26.26 (0.34) & 0.42 (0.03) \\
(A5) Activation-based re-alignment  & {\bf 40.57} (3.01) & 55.62 (0.50) & {\bf 24.07} (0.27) & 0.42 (0.03) \\
(A6) Correlation-based re-alignment~\cite{li2015convergent} & \textbf{40.57 (3.02)} & \textbf{55.63 (0.57)} & 24.36 (0.28) & 0.42 (0.03) \\
\midrule
\multicolumn{5}{c}{\textbf{B. Correlation between activations ($\uparrow$ indicates better alignment)}} \\
\midrule
(A0) No re-alignment & 0.14 (0.03) & 0.02 (0.01) & 0.01 (0.00) & 0.28 (0.08) \\
\midrule
(A1) Weight sorting~\cite{ganju2018property} & 0.29 (0.02) & 0.07 (0.01) & 0.01 (0.00)  & 0.28 (0.08)\\
(A2) Re-alignment after initialisation & 0.08 (0.01) & 0.05 (0.01) & 0.06 (0.01) & 0.28 (0.08) \\
(A3) Bottom-up weight-based re-alignment & 0.52 (0.07) & 0.31 (0.02) & 0.12 (0.01) & 0.28 (0.08) \\
(A5) Top-down weight-based re-alignment & 0.45 (0.07) & 0.25 (0.01) & 0.08 (0.00) & 0.28 (0.08) \\
(A5) Activation-based re-alignment & 0.56 (0.05) & {\bf 0.39} (0.02) & {\bf 0.20} (0.01) & 0.28 (0.08) \\
(A6) Correlation-based re-alignment~\cite{li2015convergent} & \textbf{0.57} (0.05) & 0.29 (0.02) & {\bf 0.20} (0.01) & 0.28 (0.08) \\
\bottomrule
\end{tabular}
\label{table:misalignment_scores_alignment_other_metrics}
\end{table*}

\subsection{Re-aligning ResNet architectures}\label{appendix:resnet}
ResNet architectures differ from the standard CNNs we have considered so far because they consist of blocks of convolutional layers, where a block is a sequence of layers $B=bn_2\circ conv_2 \circ relu \circ bn_1 \circ conv_1$ with a skip connection (identity or projection). 
Here, $bn$ denotes a batch normalisation layer parameterised by the mean, the variance, the weight and the bias of size equal to the number of output channels of the previous layer.
The identity skip connection means that instead of forwarding the output of the block $B(x)$ to the next block, $B(x) + x$ is forwarded instead.
The project skip connection means that instead of forwarding $B(x)$ to the next block, $B(x) + P(x)$ is forwarded, where $P(x)$ (also implemented through a convolutional layer followed by a batch normalisation layer) maps $x$ to a space of same size as $B(x)$. 

ResNet18~\cite{he2016deep} consists of one convolutional layer (followed by batch normalisation, ReLU non-linearity and max pooling), eight blocks alternating between shortcut and skip connections $B_1$ to $B_8$, an average pooling layer and an FC layer $g$ of input size 512 and output size 200 (number of classes in the Tiny-ImageNet-200 dataset).
For computational reasons (every additional re-alignment technique requires running the MIAs many times on multiple layers and feature types), and because of better re-alignment performance in the top layers which contain most of the MIA signal~\cite{nasr2019comprehensive}, we here focus on implementing the top-down weight-based re-alignment in the last layer $g$.

We start by re-aligning the top FC layer as described in Sec.~\ref{sec:aligning_layers} using the optimal permutation $\pi$, i.e., apply $\pi$ to the columns of this layer's weight matrix.
The permutation needs to be propagated to  the block below, $B_8=bn^8_2\circ conv^8_2 \circ relu \circ bn^8_1 \circ conv^8_1$.
First, we apply $\pi$ to the parameters of $bn^8_2$.
Second, we apply $\pi$ to the output channels of $conv^8_2$.
Third, to ensure consistency between $B_8(x)$ and $x$, we also apply $\pi$ to the input channels of $conv^8_1$.
Fourth and finally, as we have modified the order of input channels of $conv^8_1$, we have to propagate the permutation to the block further below, $B_7=bn^7_2\circ conv^7_2 \circ relu \circ bn^7_1 \circ conv^7_1$ and projection $P= bn_p^7 \circ conv_p^7$.
Indeed, the (summed) outputs of $B_7$ and $P$ feed into $conv^8_1$, whose input order we have modified.
Consequently, we have to apply $\pi$ to $bn^7_2$, to the output channels of $conv^7_2$, to $bn_p^7$, and to the output channels of $conv_p^7$.
The algorithm stops here: we no longer need to propagate $\pi$ to the blocks below, since we only aimed to re-align the top layer as described above.

Our re-alignment schema can be easily extended to other ResNet architectures (e.g., having more layers) and to the bottom-up direction. 
We leave this for future work.

\subsection{Meta-classifier architecture}\label{appendix:meta-classiffier}
We use a meta-classifier (MC) architecture similar to the one proposed by Nasr et al.~\cite{nasr2019comprehensive}.
It consists of multiple modules, each used to separately embed the different types of features sets in a layer, and a label embedding module.
The outputs of the modules are concatenated together and given as input to a membership classifier module.
We embed gradient features separately in each layer using a CNN architecture~\cite{nasr2019comprehensive}, consisting of a convolutional layer with kernel size equal to 100.
The CNN is followed by a dropout layer with probability 0.2 and an MLP with a hidden layer of size 128, ReLU non-linearity, and an output layer of size 64.
We embed activation features separately in each layer using the same MLP architecture as described above.
We embed each class label using a vector of learnable weights of size 16.
Finally, we use an MLP architecture as the membership classifier module.
The MLP has two hidden layers of sizes 128 and 64, each followed by ReLU non-linearity, and an output layer of size 2.

We train the MC using mini-batch gradient descent, the binary cross-entropy loss, the Adam optimiser~\cite{kingma2014adam}, and a learning rate of $\eta'=0.001$, for a maximum of 100 epochs.
We divide $\eta'$ by 2 at the end of an epoch if the validation accuracy does not improve compared to the best found so far, and stop the training when $\eta'<0.0001$.
The records used to compute the MC's validation accuracy are passed through the target model $\mathcal{M}_T$ in setup (S1) and through a \textit{validation shadow model} in setups (S2)-(S9).
The MC is thus trained on feature vectors derived from $K-1$ shadow models and early stopped based on the performance on the $K$-th shadow model.
For better performance, each mini-batch only contains features computed from a single shadow model. 
The validation shadow model is, by default, the first among the $K$ trained for all the datasets with one exception.
On Texas100, we observe a high variability between shadow models in terms of the number of completed training epochs $N_{\text{best}}$ before the best model is found.
Thus, we select the shadow model having the median value of $N_{\text{best}}$ as the validation shadow model.

We sample mini-batches differently depending on whether $D_A\cap D_{\text{target}}=\emptyset$ or $D_A = D_{\text{target}}$ (see Table~\ref{table:details} for an overview of the experiment settings).
In the first case, for every epoch we shuffle the order of records in the MC's training dataset (which is a random subset of $D_A$ of size $N_{\text{train}}$) and select, for every mini-batch, a random shadow model (reshuffling the shadow models every $K-1$ mini-batches).
An epoch consists of one pass over the $N_{\text{train}}$ records in the MC's training dataset.
Since each shadow model is trained on half of $D_A$, it follows that batches are balanced w.r.t. the membership label.
In the second case, since $D_T$ and $D_A$ have comparable sizes, if we were to use the same procedure the mini-batches would be heavily imbalanced w.r.t. the membership label. 
We thus use a different procedure: for every batch we randomly select a shadow model $\mathcal{M}_k$ and then we randomly select a balanced set of member and non-member records of $\mathcal{M}_k$'s training dataset.
We count as epoch one pass over $N_{\text{train}}$ records.

\subsection{Additional results}
\label{appendix:additional_results}
\begin{figure*}[!htbp]
\centering
\subfigure[CIFAR10]{
  \includegraphics[width=0.23\linewidth]{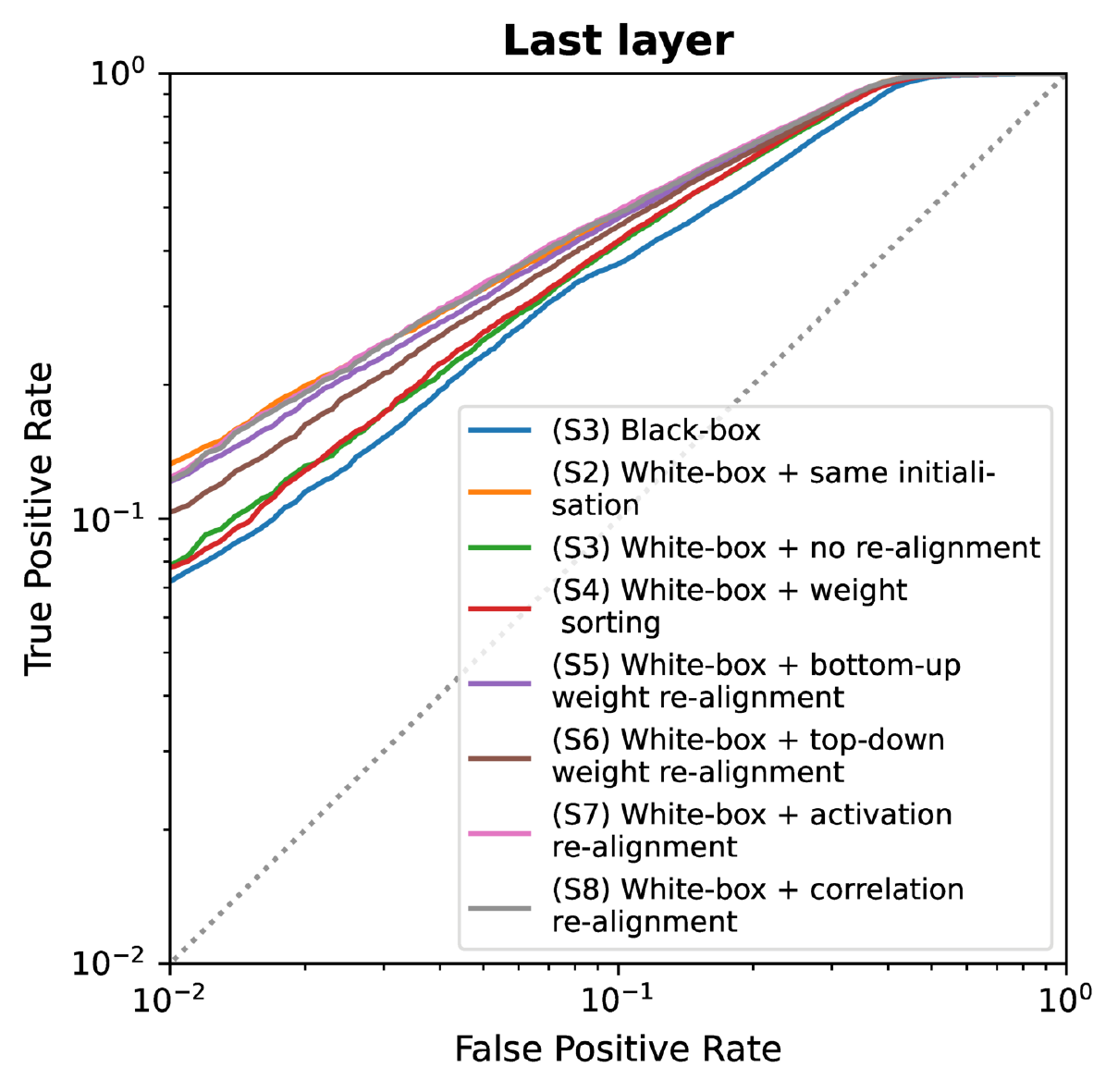}
  \includegraphics[width=0.23\linewidth]{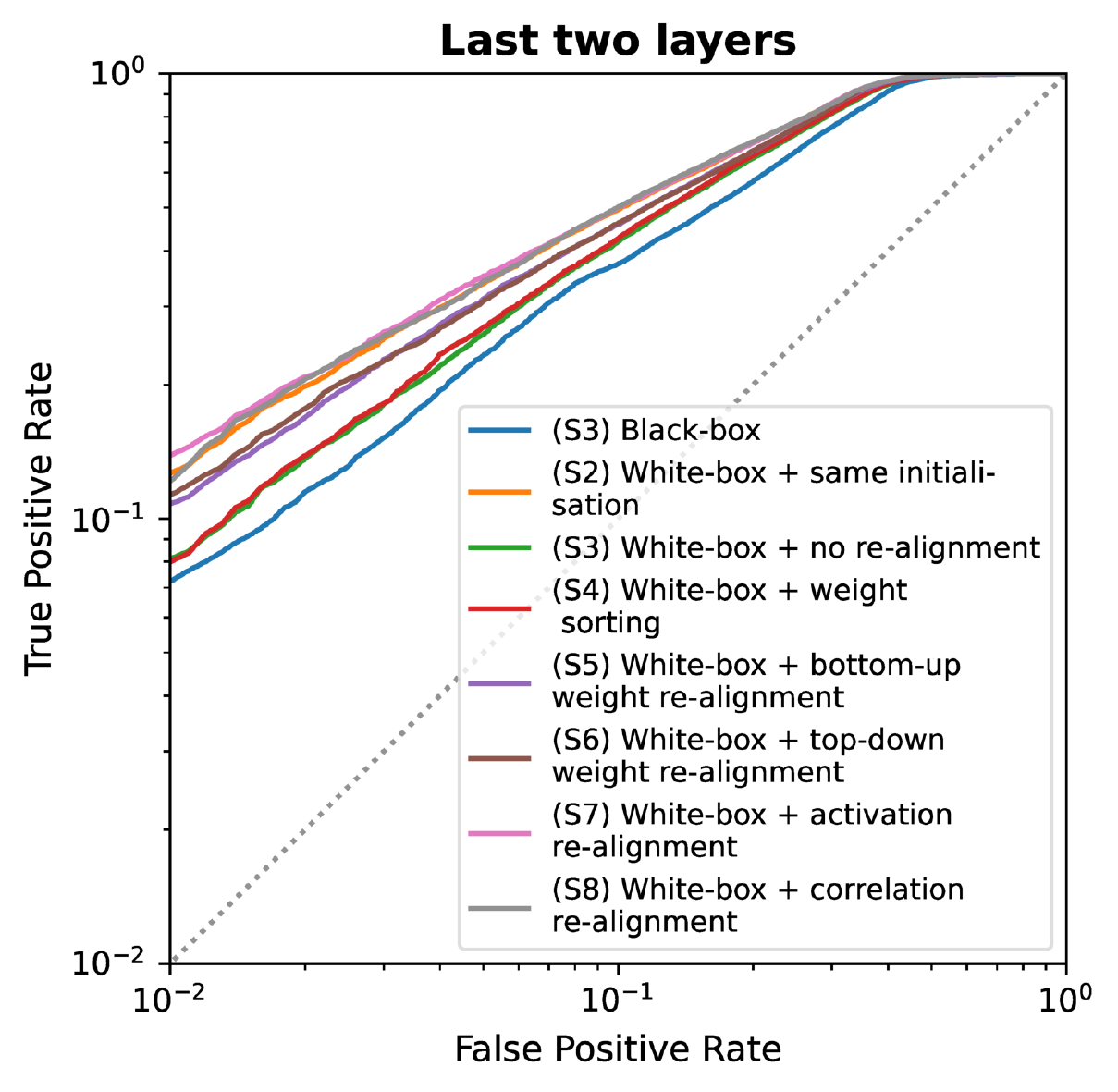}
}
\subfigure[Texas100]{
    \includegraphics[width=0.23\linewidth]{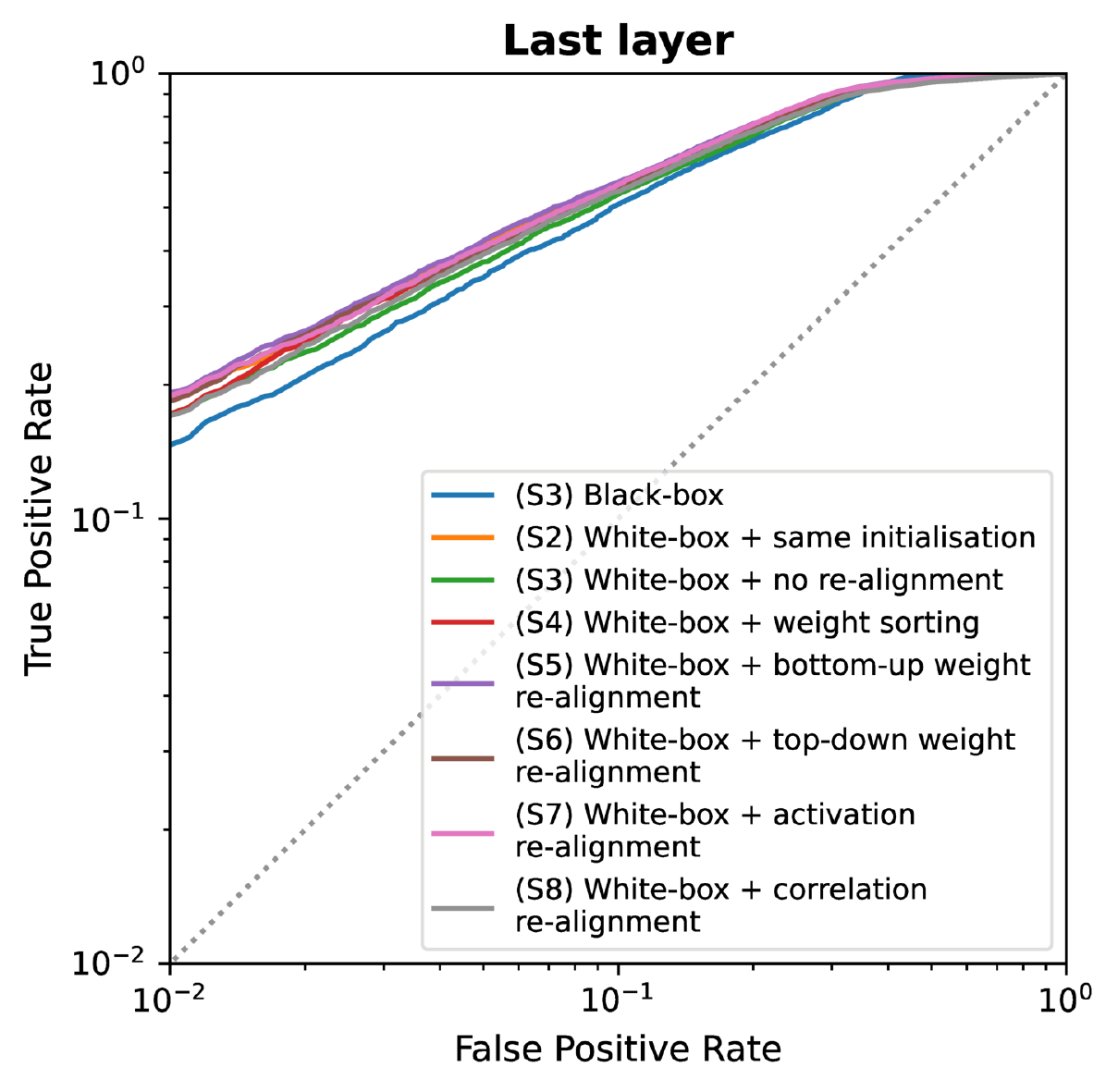}
}
\subfigure[Purchase100]{
\includegraphics[width=0.23\linewidth]{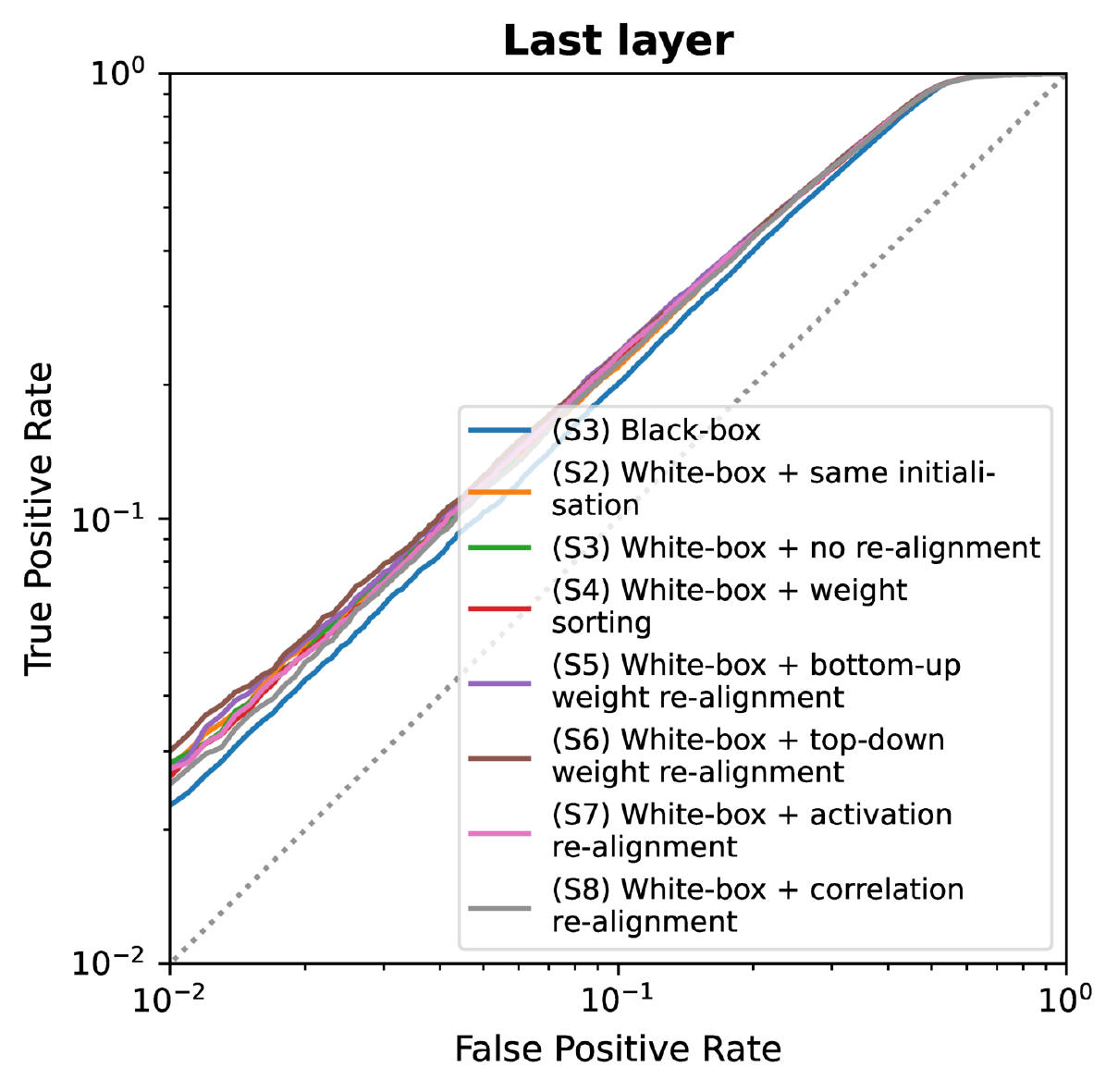}
}
\caption{\textbf{ROC curves of attacks.} The white-box MIAs use features extracted from the last layer (a-left, b, and c) or from the last two layers (a-right). We also report results for the black-box MIA using the output activations of the last layer.}
\label{fig:roc_curves}
\end{figure*}

We report additional MIA results on CIFAR10 (VGG16), Purchase 100, and Tiny-ImageNet-200 (ResNet18) datasets (Tables ~\ref{table:mia_cifar10_vgg16},~\ref{table:mia_purchase100} and ~\ref{table:mia_resnet18}, respectively) and the ROC curves on CIFAR10 (CNN4), Texas100 and Purchase 100 datasets (Table~\ref{fig:roc_curves}).

\subsection{Comparison with other baselines}\label{appendix:other-baselines}
\subsubsection{Set-based classifiers}\label{appendix:set-based}
We here describe how we extend set-based classifiers to white-box MIAs.
More specifically, we restrict to neurons in the last layer $L$, which contain most of the membership signal~\cite{nasr2019comprehensive}.
For $d=1,\ldots,D^{L-1}$, we represent how the $d^{\text{th}}$ neuron acts on a record $(x,y)$ using the following feature vector $v_d(x) \in \mathbb{R}^{D^L+3}$: (1) the output activation $x^{L-1}_d$, (2) the input activation entering the neuron of the correct label $W^L_{y,d} x^{L-1}_d$, (3) the gradients of the weights coming out the neuron $\partial \mathcal{L}/\partial W^L_{i, d} (x,y), i=1,\ldots,D^{L}$, and (4) the gradient of the bias term $\partial\mathcal{L}/\partial b^{L-1}_d(x,y)$.
We embed $v_d(x)$ using an MLP $\phi$ with one hidden layer of size 128 and an output size of 64.
We train a meta-classifier (MC) on the concatenation of the set representation $\sum_{d=1}^{D^{L-1}} \phi(v_d(x))$, the embedding of the output activations of the last layer $x^L$, and the embedding of the label $y$, with all other details the same as before.

\begin{table}[!ht]
\centering
\caption{\textbf{Results of MIAs using a set-based classifier.} 
}
\resizebox{0.48\textwidth}{!}{
\begin{tabular}{ccc}
\toprule
\textbf{CIFAR10 (CNN4)} & \textbf{Texas100 (MLP5)} & \textbf{Purchase100 (MLP4)} \\
\midrule
0.840 $\pm$ 0.014 & 0.873 $\pm$ 0.025 & 0.747 $\pm$ 0.013  \\
\bottomrule
\end{tabular}
}
\label{table:set_based_classifier}
\end{table}
\begin{table}[!ht]
\centering
\caption{\textbf{CIFAR10 (CNN4): Comparison with ~\cite{leino2020stolen}.} 
}
\begin{tabular}{l|c}
\toprule
\textbf{Method} & \textbf{MIA AUC} \\
\midrule
 Bayes-WB (Leino et al.~\cite{leino2020stolen}) & 0.708 $\pm$ 0.005 \\
IA only, not re-aligned (S3)  & 0.726 $\pm$ 0.006 \\
IA only, re-aligned (S6) & \textbf{0.805} $\pm$ 0.014 \\
\bottomrule
\end{tabular}
\label{table:comparison_stolen_memories}
\end{table}
\begin{table}[!ht]
\centering
\caption{\textbf{Comparison between black-box attacks.} 
}
\resizebox{0.48\textwidth}{!}{
\begin{tabular}{lccc}
\toprule
& \textbf{CIFAR10} & \textbf{Texas100} & \textbf{Purchase100} \\
\midrule
(S2) Shokri et al.~\cite{shokri2017membership} & 0.810$\pm$0.011 & 0.843$\pm$0.023  & 0.736$\pm$0.010  \\
(S2) LiRA~\cite{carlini2022membership} & 0.815$\pm$0.017 & 0.835$\pm$0.059 & 0.734$\pm$0.016 \\
(S3) Shokri et al.~\cite{shokri2017membership} & 0.815$\pm$0.011 & 0.859$\pm$0.015 & 0.733$\pm$0.012 \\
(S3) LiRA~\cite{carlini2022membership} & 0.803$\pm$0.018 & 0.829$\pm$0.055 & 0.731$\pm$0.015 \\
\bottomrule
\end{tabular}
}
\label{table:black-box-mias}
\end{table}

\subsubsection{White-box MIA of Leino and Fredrikson~\cite{leino2020stolen}}\label{appendix:leino}
We compare our approach with the Bayes-WB approach of Leino and Fredrikson~\cite{leino2020stolen}.
This approach trains proxy models whose goal is to replicate the ``semantic meaning'' of the target model features in an internal layer $l$.
To this end, the proxy model is initialised using the target model weights in all the layers up to $l$.
These layers are then frozen, while the remaining layers are trained from scratch on shadow data.
Then, for each target record $(x, y)$, the method computes local linear approximations $(W^x_T, b^x_T), (W^x_1, B^x_1)$ of how the remaining layers restricted to the output class $y$ act on the input $x$, in the target and proxy model, respectively.
Finally, a Bayesian approach exploiting the differences between the behaviour of the linear approximations on the target record is used to produce a membership score: $sigmoid((W^x_T - W^x_1)^t x + (b^x_T - b^x_1))$.
This is shown to be optimal on binary logistic regression under specific assumptions on the data distribution $\mathcal{D}$.

We re-implement their approach using the last layer $l=L$, which is linear, meaning that we can use the pure signal, without the need to compute local approximations.
Indeed, we were unable to replicate the influence-based linear approximators.
In this scenario, the Bayes-WB approach uses as features the IAs.
We train $K=10$ proxy models and use their average weights as recommended by the authors.
For a fair comparison, we also run our alignment-based attack using the same features. 
To minimise the alignment error in the top layer, we choose top-down weight matching (S6).

\subsubsection{Black-box LiRA attack~\cite{carlini2022membership}}\label{appendix:lira}
We present results of the black-box LiRA attack proposed by Carlini et al.~\cite{carlini2022membership}. More specifically, we implement the offline LiRA attack using the same number of shadow models as our attacks. 
We perform a fair comparison with Shokri et al.'s method~\cite{shokri2017membership} on the same datasets and models under scenarios (S2) and (S3). We report the mean Area Under the Curve over 10 repetitions with 95\% confidence interval. 
Our results show that the two approaches perform on par. We attribute this to the small number of shadow models used ($K=10$), as the LiRA method requires hundreds of shadow models to achieve its best performance~\cite{carlini2022membership}.
For computational reasons, we cannot in this work train hundreds of shadow models.
However, if we could, fairly evaluating white-box MIAs against LIRA would also require adapting the former to the targeted setting, by learning a record-dependent decision boundary.
We leave this very interesting idea for future work, as discussed in Sec.~\ref{sec:discussion}.

\subsection{Defenses}\label{appendix:defenses}
We here describe our experiments with models trained using defenses: regularisation and differential privacy~\cite{dwork2006calibrating}.
First, we train the same VGG16 network as before, on CIFAR10, using the largest weight decay which still allows the model to converge ($3\times10^{-3}$ as opposed to $5\times10^{-4}$ in the previous experiments). 
As upon convergence  this model is highly overfitted (a known source of MIA vulnerability~\cite{yeom2018privacy}), we stop the training when the gap between validation and train accuracy exceeds 5\%, at a small cost in utility (78.4\% instead of 82.6\%).
We observe no gap between black-box and white-box attacks in the (S1) setting (0.531 vs 0.535), suggesting that there would be no benefit in re-alignment and that this model is overall robust to MIAs.
Second, we trained VGG16 using the DP-SGD algorithm implemented in the opacus library~\cite{yousefpour2021opacus}, however values of $\epsilon\leq100$ lead to very low model accuracy.
We switched to the smaller CNN4 network and tried the recommended small epsilon values, which gave poor utility. For completeness, we trained a model using $\epsilon=50$ until it converged to 49\% (48.6\%) test (train) accuracy.
Such large $\epsilon$ values can still provide empirical privacy protection as suggested by~\cite{jayaraman2019evaluating}.
We indeed found this network to be robust to both black-box and white-box MIAs, which achieve an AUC of 0.506.
Third, we train an MLP on 50000 samples of Purchase100 (instead of 20000 as before) using $\epsilon=50$ to achieve 69.2\% (72.1\%) test (train) accuracy.
As before, both black-box and white-box attacks are mitigated as they achieve close to random performance (0.504).
We conclude that implementing these defenses effectively reduces the risk of MIAs albeit at a cost in accuracy.

\begin{table*}[!ht]
\centering
\caption{\textbf{CIFAR10 (VGG16): Results of white-box MIAs using features from multiple layers.} 
}
\begin{tabular}{l|c|ccc}
\toprule
\multirow{2}{*}{\textbf{Source of meta-classifier training features}} & \textbf{Black-box} & \multicolumn{2}{c}{\textbf{White-box}} \\
\cmidrule{2-4}
& Last layer  & Last layer & Last two layers \\
\midrule
(S1) Target model (auditor)  & 0.637 $\pm$ 0.013  & 0.686 $\pm$ 0.005 & 0.689 $\pm$ 0.009 \\
\midrule
(S2) Shadow models (same WI) & 0.640 $\pm$ 0.009 & 0.689 $\pm$ 0.004 & 0.691 $\pm$ 0.005 \\
\midrule
(S3) Shadow models (all seeds $\neq$)  & 0.630 $\pm$ 0.013 & 0.678 $\pm$ 0.006 & 0.679 $\pm$ 0.005 \\
(S5) Shadow models (all seeds $\neq$) + bottom-up weight re-al.  & - & 0.672 $\pm$ 0.015 & 0.686 $\pm$ 0.005 \\
(S6) Shadow models (all seeds $\neq$) + top-down weight re-al. & - & 0.681 $\pm$ 0.008 & 0.680 $\pm$ 0.011 \\
(S7) Shadow models (all seeds $\neq$) + activation re-al.  & - &  0.684 $\pm$ 0.008 & 0.686 $\pm$ 0.009 \\
\bottomrule
\end{tabular}
\label{table:mia_cifar10_vgg16}
\end{table*}

\begin{table*}[!ht]
\centering
\caption{\textbf{Purchase100 (MLP4): Results of MIAs using different features of the last layer.} 
}
\begin{tabular}{l|c|c|c}
\toprule
\textbf{Source of meta-classifier training features} & \textbf{Black-box (OA)} & \textbf{White-box (OA + IA + G)} & \textbf{White-box (G)}\\
\midrule
(S1) Target model (auditor) & 0.733 $\pm$ 0.010 & 0.757 $\pm$ 0.013 & 0.734 $\pm$ 0.013 \\
\midrule
(S2) Shadow models (same WI) & 0.736 $\pm$ 0.010 & 0.745  $\pm$ 0.012 & 0.729 $\pm$ 0.013 \\
\midrule
(S3) Shadow models (all seeds $\neq$) & 0.733 $\pm$ 0.012 & 0.746 $\pm$ 0.011 & 0.727 $\pm$ 0.014 \\
(S4) Shadow models (all seeds $\neq$) + weight sorting~\cite{ganju2018property} & -  & 0.745 $\pm$ 0.012 & 0.724 $\pm$ 0.014 \\
(S5) Shadow models (all seeds $\neq$) + bottom-up weight re-al. & - & 0.750 $\pm$ 0.013 & 0.726 $\pm$ 0.014 \\
(S6) Shadow models (all seeds $\neq$) + top-down weight re-al. & - & 0.749 $\pm$ 0.013 & 0.730 $\pm$ 0.013 \\
(S7) Shadow models (all seeds $\neq$) + activation re-al. & - & 0.746 $\pm$ 0.014 & 0.730 $\pm$ 0.014 \\
(S8) Shadow models (all seeds $\neq$) + correlation re-al. & - & 0.744 $\pm$ 0.013 & 0.729 $\pm$ 0.014\\
\bottomrule
\end{tabular}
\label{table:mia_purchase100}
\end{table*}

\begin{table*}[!ht]
\centering
\caption{\textbf{Tiny-ImageNet-200 (ResNet18): Results of MIAs using different features of the layset layer.} 
}
\begin{tabular}{l|c|ccc}
\toprule
\multirow{2}{*}{\textbf{Source of meta-classifier training features}} & \textbf{Black-box} & \multicolumn{3}{c}{\textbf{White-box}} \\
\cmidrule{2-5}
& \textbf{OA}  & \textbf{OA + IA + G} & \textbf{G} & \textbf{IA} \\
\midrule
(S1) Target model (auditor) & 0.597 $\pm$ 0.010 &  0.693 $\pm$ 0.008 & 0.669 $\pm$ 0.005 & 0.688 $\pm$ 0.012  \\
\midrule
(S2) Shadow models (same WI) & 0.586 $\pm$ 0.002 &  0.694 $\pm$ 0.005 & 0.669 $\pm$ 0.007 & 0.682 $\pm$ 0.010 \\
\midrule
(S3) Shadow models (all seeds $\neq$) & 0.585 $\pm$ 0.002 & 0.689 $\pm$ 0.007 & 0.671 $\pm$ 0.008  & 0.667 $\pm$ 0.002 \\
(S7) Shadow models (all seeds $\neq$) + top-down weight re-al. & - & 0.691 $\pm$ 0.008 & 0.670 $\pm$ 0.006 & 0.675 $\pm$ 0.009 \\
\bottomrule
\end{tabular}
\label{table:mia_resnet18}
\end{table*}

